\documentclass[final,authoryear,5p,times,twocolumn]{elsarticle}

\usepackage{amssymb}
\usepackage{fixltx2e}

\usepackage{color}

\journal{Nuclear Instruments and Methods in Physics Research Section A}

\begin{document}

\begin{frontmatter}



\title{Analysis techniques and performance of the Domino Ring Sampler version 4 based readout for the MAGIC telescopes}


\author[adres1]{Julian Sitarek}
\ead{jsitarek@ifae.es}
\address[adres1]{IFAE, Edifici Cn., Campus UAB, E-08193 Bellaterra, Spain}

\author[adres3a,adres3b]{Markus Gaug}
\ead{Markus.Gaug@uab.cat}
\address[adres3a]{F\'isica de les Radiacions, Departament de F\'isica, Universitat Aut\`onoma de Barcelona, 08193 Bellaterra, Spain.}
\address[adres3b]{CERES, Universitat Aut\`onoma de Barcelona-IEEC, 08193 Bellaterra, Spain.}

\author[adres4]{Daniel Mazin}
\ead{mazin@mpp.mpg.de}
\address[adres4]{Max-Planck-Institut f\"ur Physik, D-80805 M\"unchen, Germany}

\author[adres5]{Riccardo Paoletti}
\ead{paoletti@pi.infn.it}
\address[adres5]{Universit\`a  di Siena, and INFN Pisa, I-53100 Siena, Italy}

\author[adres6]{Diego Tescaro}
\ead{diegot@ifae.es}
\address[adres6]{Inst. de Astrof\'{\i}sica de Canarias, E-38200 La Laguna, Tenerife, Spain}

\begin{abstract}
Recently the readout of the MAGIC telescopes has been upgraded to a new system based on the Domino Ring Sampler version 4 chip.
We present the analysis techniques and the signal extraction performance studies of this system. 
We study the behaviour of the baseline, the noise, the cross-talk, the linearity and the time resolution.
We investigate also the optimal signal extraction. 
In addition we show some of the analysis techniques specific to the readout based on the Domino Ring Sampler version 2 chip, previously used in the MAGIC~II telescope.
\end{abstract}

\begin{keyword}
Gamma-ray astronomy, Cherenkov telescopes, signal extraction
\end{keyword}

\end{frontmatter}


\section{Introduction}\label{sec:intro}
MAGIC (Major Atmospheric Gamma Imaging Cherenkov, \citet{magic2_status, cortina2011}) is a system of two imaging atmospheric Cherenkov telescopes (IACTs, \citet{whipple}), each equipped with a mirror dish  of 17$\,$m diameter. 
MAGIC is located on the Canary Island of La Palma at the height of 2200 m a.s.l. 
The telescopes are built to measure gamma rays in the energy range from about 50 GeV to 50 TeV by detection of short and weak Cherenkov light flashes from the extensive air showers.
This requires fast time response and high sampling rate of the system in order to decrease the exposure to noise.

Presently both telescopes are using a readout based on the Domino Ring Sampler version 4 chip\footnote{http://drs.web.psi.ch/} (DRS4)~\citep{ritt_drs4, bpt13} operated at a sampling speed of 2 GSamples/s.
Such sampling speeds, larger than in other IACTs~\citep{hess_fadc,veritas_fadc}, are used in MAGIC since 2007, and allow us to exploit the timing information in recordered showers. 
Another presently operating IACT using the DRS4 chip in its readout, is FACT~\citep{fact}.
In addition, the DRS4 is the heart of the Dragon system \citep{cta_drs4}, one of the possible readout systems considered for the next generation major IACT project called Cherenkov Telescope Array (CTA,~\citealp{cta}).
Among the other analogue memories which are considered for the readout of CTA are TARGET \citep{target} and NECTAr \citep{nectar}.

In this paper, we describe the pre-processing analysis procedures used to extract and calibrate the signal in each channel of the MAGIC telescopes. 
We also evaluate the performance of the basic parameters of the signal extraction. 
In addition, we describe methods used in the analysis of the data taken with the previous readout (based on the DRS2 chip, \citealp{ritt_drs2, magic_drs2_2004, bitossi}) of the MAGIC~II telescope. 
The older readout systems of the MAGIC~I telescope are described in detail in~\citet{magic_fadc, magic_mux}.

\section{The MAGIC telescopes}
Very High Energy (VHE, $\gtrsim 100\,$GeV) gamma rays entering the atmosphere produce cascades of secondary particles. 
The charged particles moving faster than the propagation speed of light in the atmosphere will produce very short (of the order of $1\,$ns) flashes of optical and UV light (the so-called Cherenkov radiation, \citet{cherenkov}). 
This light is emitted in a narrow cone with small half-opening angles of $\lesssim 1^\circ$ and illuminates an area on the ground with a radius of about $120\,$m. 

\begin{figure}
\includegraphics[width=9truecm]{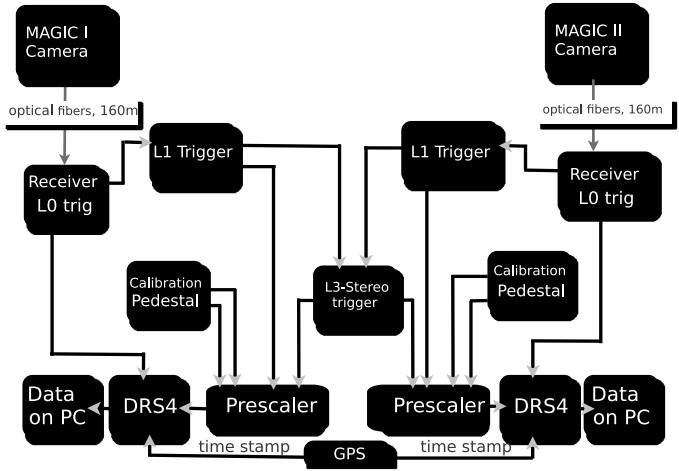}
\caption{
Electronic chain of the MAGIC telescopes.
}
\label{fig:chain}
\end{figure}

IACTs, such as the MAGIC telescopes, can focus the Cherenkov light onto the camera composed of the order of $1000$ individual photodetectors (pixels). 
Until summer 2012 the cameras of the MAGIC~I and MAGIC~II telescopes were equipped with 577 and 1039 photomultipliers (PMTs), respectively. 
At that time the  MAGIC~I camera was equipped with two kinds of pixels: 397 smaller, inner pixels with a FoV of $0.1^\circ$, and 180 larger, outer pixels with a FoV of $0.2^\circ$ \citep{cortina2005}.
The MAGIC~II camera is composed only of smaller pixels.
The full electronic chain of the MAGIC telescopes is schematically shown in Fig.~\ref{fig:chain}.
The electrical signal from each pixel is converted into an optical one using a vertical-cavity surface emitting laser (VCSEL) diode in the camera \citep{jordi}.
Afterwards it is transmitted, still as an analog signal, via an optical fiber to the control house hosting the readout electronics.
The signal then arrives to the so-called receivers where it is converted back into an electrical pulse and split into a trigger and readout branches. 
Between February 2007 and June 2011, the MAGIC~I readout was based on optical multiplexer and off-the-shelf FADCs~\citep{magic_mux}.
The second MAGIC telescope, in operation since 2009, was first equipped with the Domino Ring Sampler version 2 chip (DRS2,~\citealp{magic_drs2}). 
During the first stage of the MAGIC upgrade in 2011, both telescopes were equipped with a readout based on the DRS4 chip \citep{bpt13}.
In a second stage of the upgrade, the camera of MAGIC~I has been replaced by a close copy of the one presently installed in MAGIC~II. 
As bigger outer pixels covered only the outer part of one of the MAGIC cameras and in the present system there are only small pixels, in this paper we concentrate on those.

The DRS4 readout system is based on an array of 1024 capacitors for each channel.
When running the system with a sampling speed of 2 GSamples/s, the input signal is stored in the analog form in the capacitors with a switching period of 500 ps, which results in a 512 ns deep buffer.
After a trigger occurs, the sampling is stopped and the charges of the capacitors are read out by an ADC of 14bit precision at a speed of 32 MHz \citep{bpt13}. 
The studies presented in this paper are based on data in which the waveforms for a time span of 40 ns (80 samples) around the pulse position (the so-called region of interest, RoI) were stored for each event and each pixel.
Very recently the number of saved samples have been reduced to 60. 
With this sampling range, even large showers with long time development are contained in the readout time window. 
Such a readout window also ensures that the pulses will not be truncated due to the jitter or drifts of the trigger signal.

The calibration of the readout signals is done using calibration laser pulses of a wavelength of 355nm and with a Full Width Half Maximum (FWHM) of $\sim1.1\,$ns that illuminate homogeneously the entire camera.
The intensity of the light in the calibration pulses can be set to various values, spanning the whole dynamic range of the readout.
Each stored event is tagged with a time stamp from a rubidium clock synchronized with a GPS system.

\section{Signal processing}\label{sec:signal_processing}
The purpose of the pre-processing analysis is to obtain two pieces of information for each pixel in a given event: the total signal (charge) and the arrival time.
The signal is converted from the integrated readout counts (i.e. summed up ADC counts from 6 consecutive time samples) to photoelectrons (phe) according to the F-Factor (excess noise factor) method~\citep[see e.g.][]{ffactor, gbcr05}. 
For the presently used integration window of 6 time samples (i.e. $3\,$ns), the conversion factor is typically $\sim90$ readout counts per phe.

A single photoelectron generates a signal with an amplitude of the order of $30$ readout counts.
However, it should be noted that the individual photoelectrons come at slightly different times, both due to the time spread in the PMT and due to the intrinsic time spread of the calibration/cherenkov light flashes.
By scaling down a $\sim 100\,$phe pulse which includes all those time spreads we obtain an effective photoelectron which is broader and has the amplitude of $\sim18$ readout counts. 

The full span of 14 bit ADC used in the readout is $2\,$V, thus one readout count corresponds to 122$\,\mathrm{\mu V}$ output voltage.
However as the DRS4 has a differential gain of 2, one readout count corresponds to $\sim 60\,\mathrm{\mu V}$ at the board input.

The position of the integration window is adjusted for each pulse such that it maximizes the obtained signal over the whole readout window (the so-called ``sliding window'' method).
For each event, we select the pixels, which are likely to contain information about the shower based on their signals and arrival times in the so-called time image cleaning procedure \citep{magic_time}.
The individual pixel charges are later used in the parametrization of the shower images~\citep{hillas85}. 
In addition, timing parameters, if determined precisely enough, can be used to further enhance the performance of the telescopes e.g. in gamma/hadron discrimination~\citep{magic_time, magic_advanced}.

All the procedures described in this paper are included in the standard analysis software for the MAGIC telescopes (MARS,~\citet{magic_mars}) and used in the automatic data processing chain.
Moreover, some of them (e.g. the baseline correction) are done online during data taking by the MAGIC data acquisition program. 

\subsection{Baseline with triggers arriving at fixed time intervals} \label{sec:baseline}

\begin{figure}[t]
\includegraphics[width=9truecm]{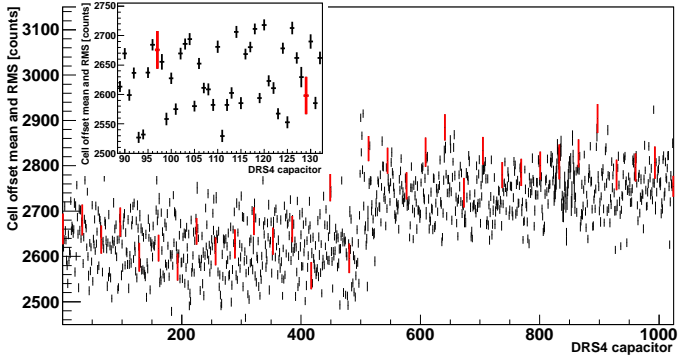}
\caption{
Cell offsets of 1024 individual capacitors of one channel of the DRS4 chip, operating in the RoI mode. 
Vertical error bars show the standard deviation of the offset value for a given capacitor. 
Every 32$^\mathrm{nd}$ capacitor is marked with thick line in red color.
The inside panel zooms into some of the capacitors.
}
\label{fig:baseline}
\end{figure}

In Fig.~\ref{fig:baseline} we show the mean cell offset (baseline) and its RMS, as a function of the absolute position of the capacitor in the domino ring for a typical DRS4 channel.
Each capacitor of each DRS4 channel has its own cell offset. 
The differences in the mean offset caused by the physical differences in the storage cells are much larger than the RMS of the baseline of the individual cells. 
Also due to the internal construction of the DRS4, there is a relatively large step in the baseline in the middle of the readout ring (i.e. at capacitor position 512).
Thus, the offset of the individual capacitors has to be calibrated in order to assure low electronic noise of the readout. 
In principle, one can apply a relatively simple calibration by computing the mean offset as a function of the capacitor number, using dedicated pedestal runs. 
This mean offset can then be subtracted from the signal in a given capacitor. 

However, as one can see in Fig.~\ref{fig:baseline}, in the RoI operation mode every 32$^\mathrm{nd}$ capacitor in the DRS4 ring has a larger RMS than the rest of the capacitors. 
Also its two neighbouring capacitors are partially affected. 
Note that as the triggers can occur at arbitrary times their relative position in the RoI will differ from one event to another. 
In fact the offset value of those capacitors depends on the relative position of the capacitor in the RoI. 
Such an effect, if not taken into account, would produce a small spike in every 32$^\mathrm{nd}$ capacitor of the readout. 
In principle one can interpolate every 32$^\mathrm{nd}$ capacitor, but this would result in loss of about 3\% of the total pulse shape information, which we found unacceptable.  
Therefore, we perform an improved calibration by computing the mean offset as a function of both the absolute position in the domino ring (1024 individual capacitors) 
and its relative position (80 possibilities in the RoI80, mode of operation).
Such a procedure removes nearly all of those spikes.
As a result also the fluctuations of the pedestal are reduced (the RMS of the baseline obtained from values of 80 capacitors in each event is reduced from $\sim 8$ to $\sim 7.5$ counts for the readout stand-alone noise).
The offset values for such a pedestal calibration are computed using a dedicated $\sim 10\,$min run taken before each observation night. 
We observed that a change of the temperature of the receivers and the readout electronics of about 2 degrees can change the baseline level by about 6 readout counts.
As the temperature of the readout boards is maintained constant within $\sim 1\,$degree C during the data taking, the baseline calibration normally does not have to be repeated during the observations.
The above described calibration is performed online by the data aquisition software before storing the data on the disk.
During the data taking we take additional interleaved pedestal events to correct for any slow drifts of the baseline.

\subsection{DRS4 time pedestal correction with randomly arriving triggers}

\begin{figure}[t]
\includegraphics[width=9truecm]{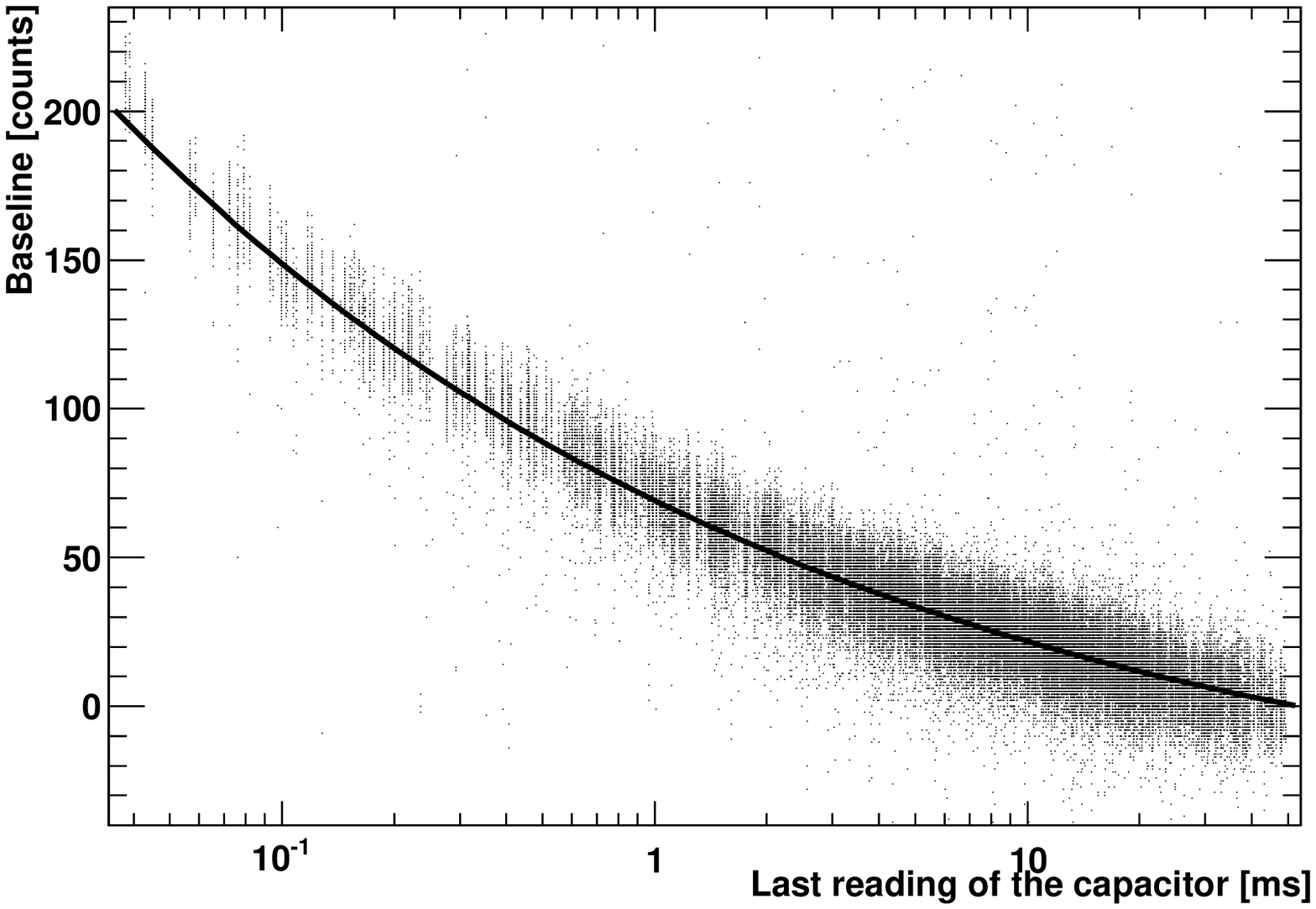}\\
\includegraphics[width=9truecm]{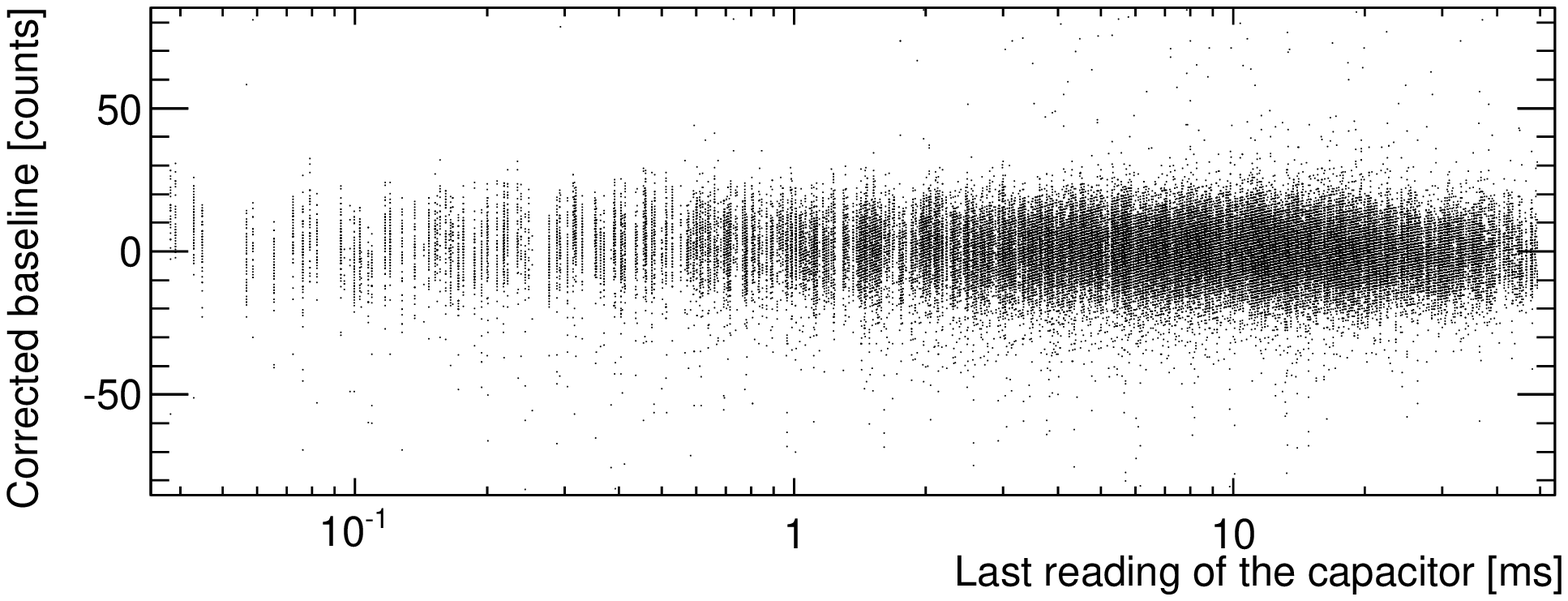}
\caption{
The dependence of the offset of individual capacitors on the time lapse to the previous reading of this capacitor for a typical DRS4 channel. 
The thick solid line shows the power-law function used for the correction. 
Top panel: raw data before the correction, bottom panel: data after correction
}
\label{fig:drs_timeped}
\end{figure}

\begin{figure}
\centering
\includegraphics[width=4.3truecm]{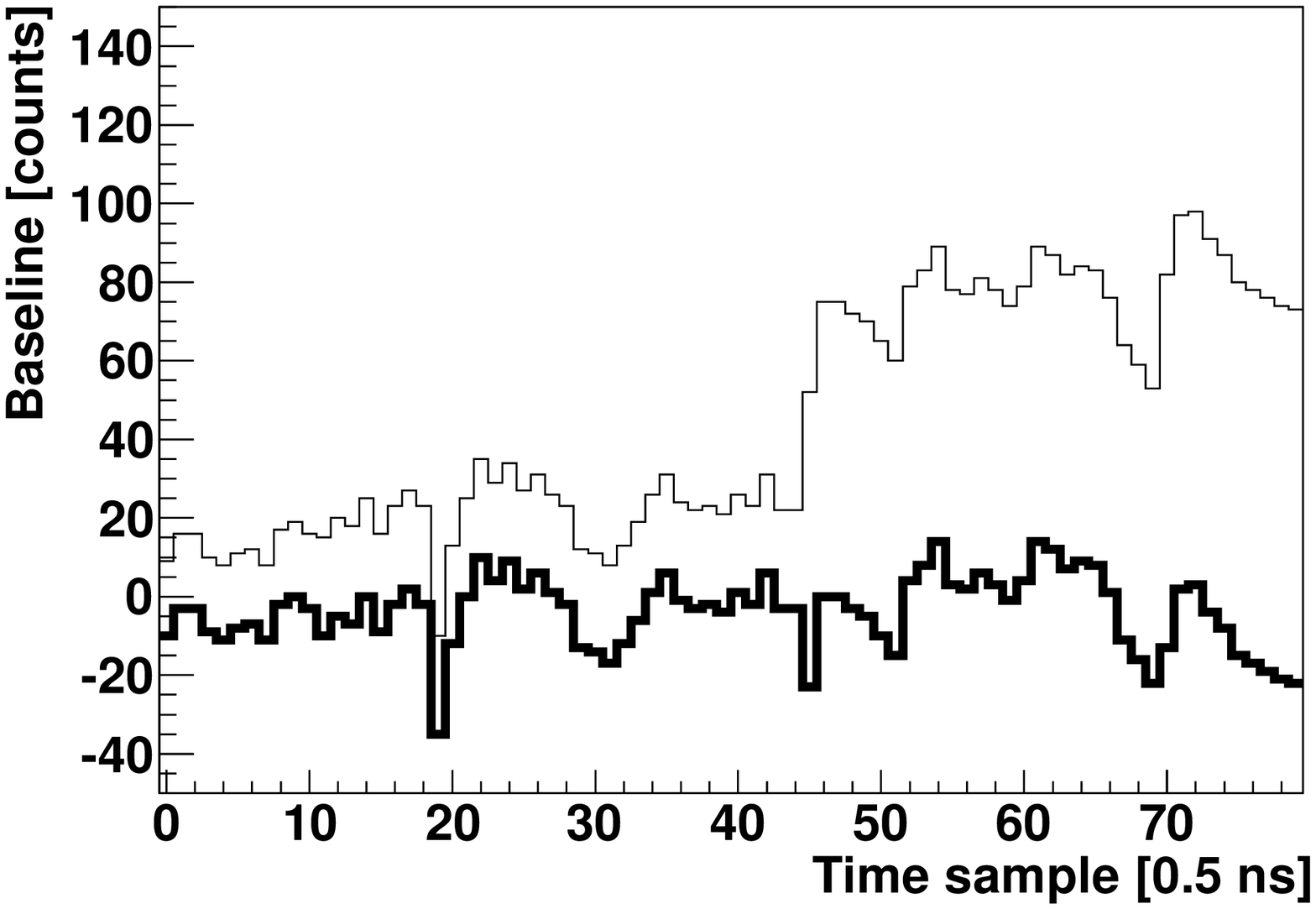}
\includegraphics[width=4.3truecm]{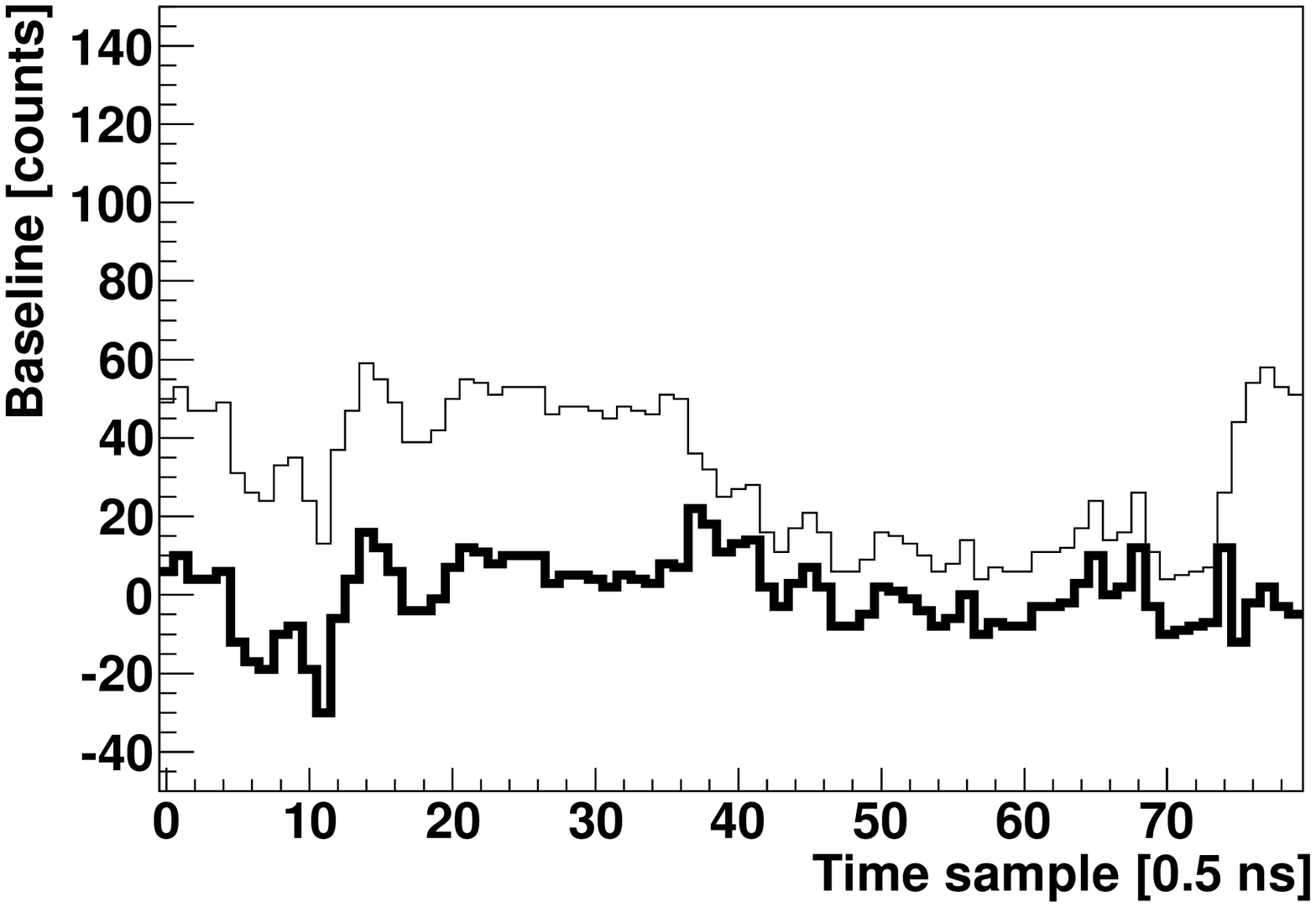}\\
\includegraphics[width=4.3truecm]{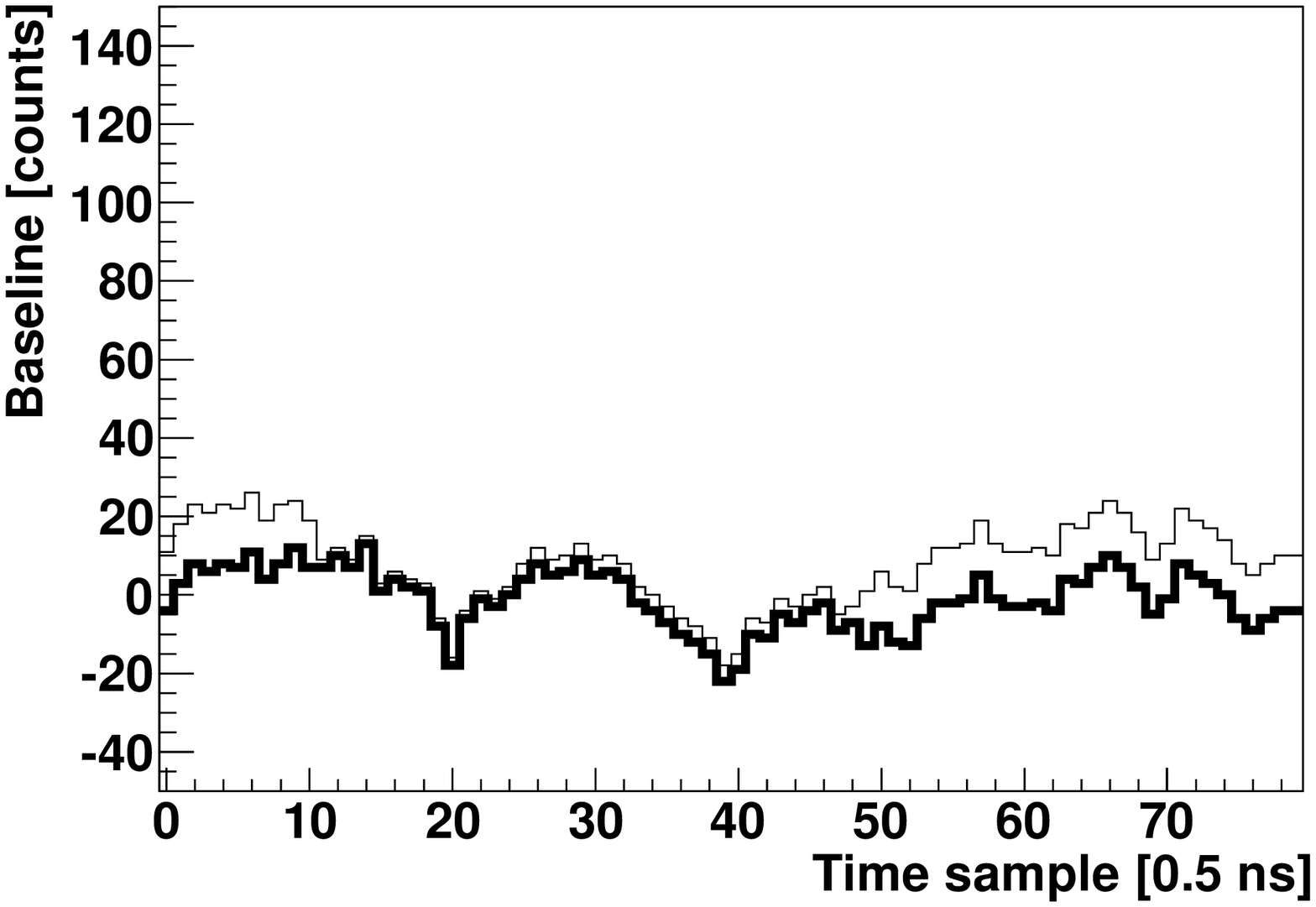}
\includegraphics[width=4.3truecm]{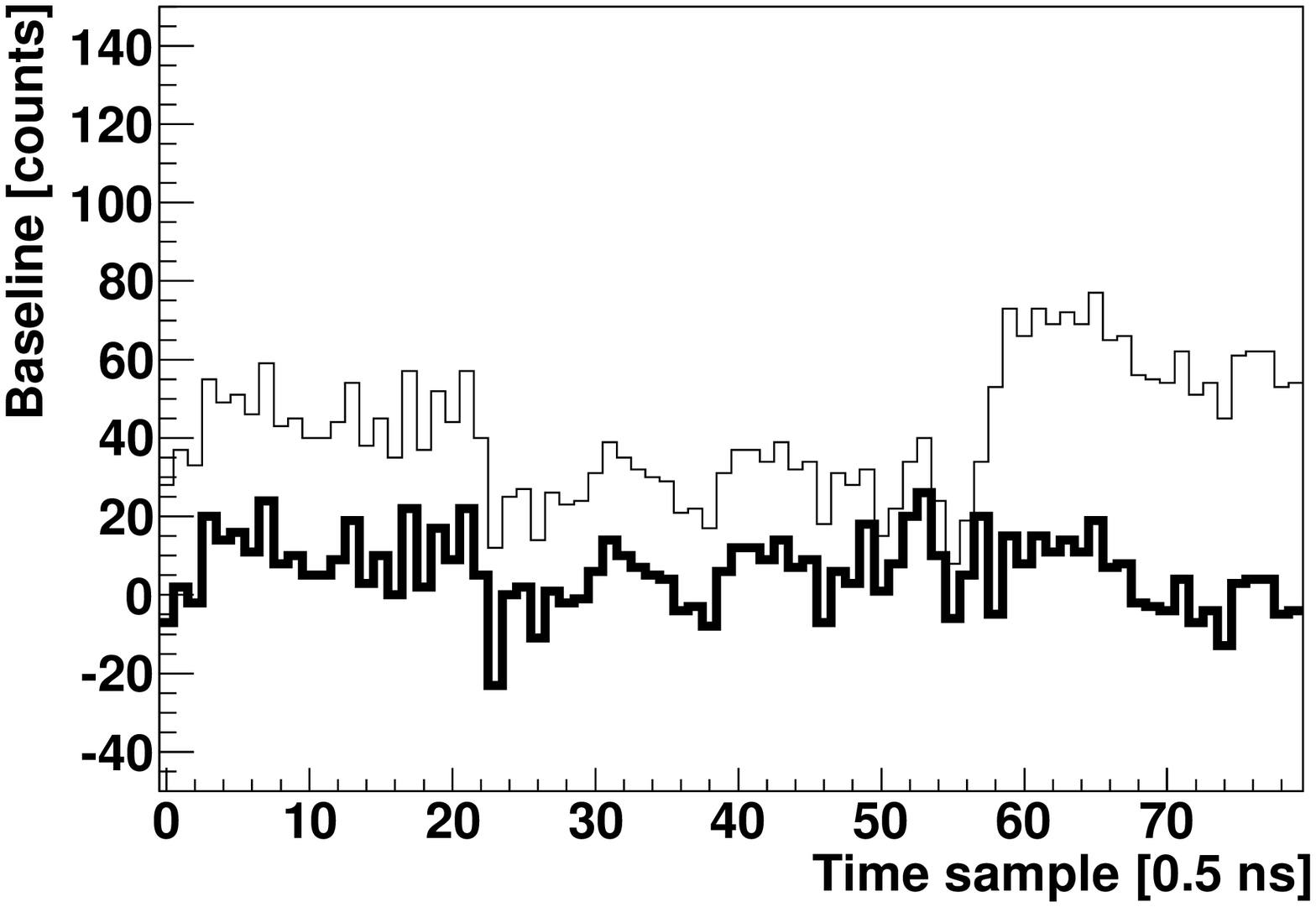}

\caption{
Four examples of baseline jumps due to the pedestal time dependence effect. 
The thin lines show the original baseline and the thick lines the corrected one. 
An amplitude of $1\,$phe signal corresponds to $\sim 30\,$counts.
}
\label{fig:pedjumps}
\end{figure}

Even after applying the calibration described in Section~\ref{sec:baseline}, the baseline of the DRS4 channel is not fully stable for randomly arriving triggers.
In fact, the DRS4 chip exhibits a dependence of the baseline on the time lapse to the last reading of a given capacitor.
Since for each event the DRS4 stops at a different part of the ring and only a limited number of capacitors are read out, it may happen that a part of the capacitors of the current readout region has been read out more recently than the rest.
If not corrected, such an effect would produce a step in the baseline, even within one event when a discontinuity in the time readout lapse lies within the current readout window. 
Typically the time between two consecutive readings of the same capacitor is of the order of $10\,$ms, but if two events trigger very fast one after another, 
and happen to occur in the same part of the ring, it may go down to few tens of microseconds.
The time lapse dependence of the cell offset can produce a jump in the baseline up to 200 counts (equivalent to an amplitude of $\sim7\,$phe). 
This time lapse dependence of the baseline is very similar (within a few \%) for all the DRS4 chips and can be fitted with a simple power-law function (see Fig.~\ref{fig:drs_timeped}).
This power-law function is used to correct the cell offsets (see Fig.~\ref{fig:pedjumps}). 
After applying this correction, the steps in the baseline of the DRS4 channels disappear. 
Apart from these effects, the baseline remains stable within 1-2 counts at the time scale of an hour, and can be estimated either with a dedicated pedestal run (taken before the data taking), or with interleaved pedestal events (taken at regular time intervals during data taking).
For normal observations this calibration is performed online by the data aquisition software before storing the data on disk.

Compared to that, the DRS2 baseline was more unstable.
It showed jumps of the baseline of tens of counts between consecutive events and had to be estimated on event-by-event basis from the waveform in the same event. 
The first 16 time samples of the readout window, which normally do not contain signals, neither from the calibration pulses nor from the cosmic showers, had been used for this baseline estimation. 
The need for such a procedure naturally increases the variation of the reconstructed signal. 
Assuming non-correlated noise in consecutive time samples, the value of the baseline can be estimated to a precision of $baselineRMS / \sqrt{16}$.  
After subtracting this baseline the noise in an 8-samples window increases by a factor of $\sqrt{1+8/16} = 1.22$ with respect to the case of a perfectly constant and known baseline. 

\subsection{Signal extraction}

\begin{figure}[t]
\includegraphics[width=9truecm]{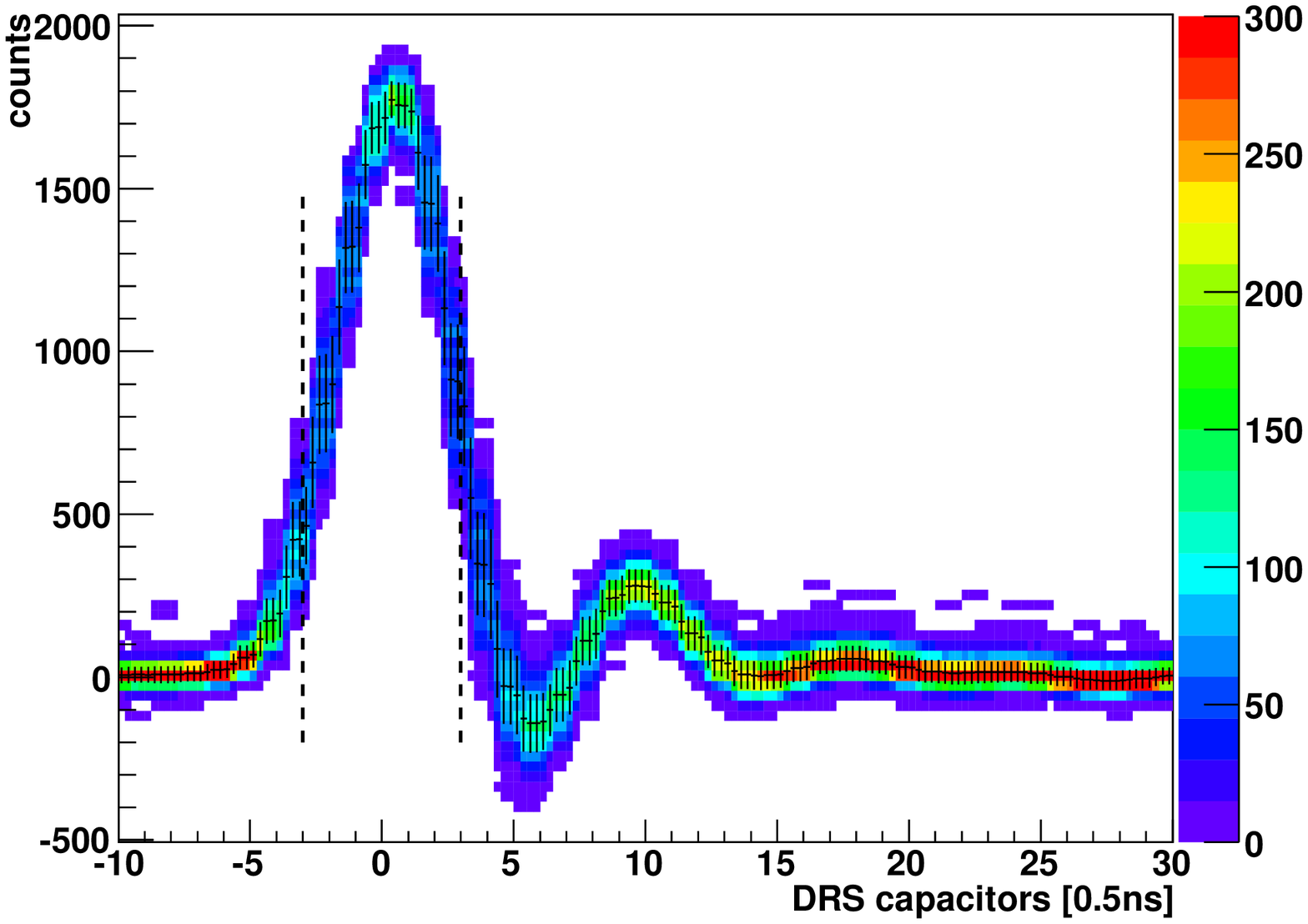}\\
\includegraphics[width=9truecm]{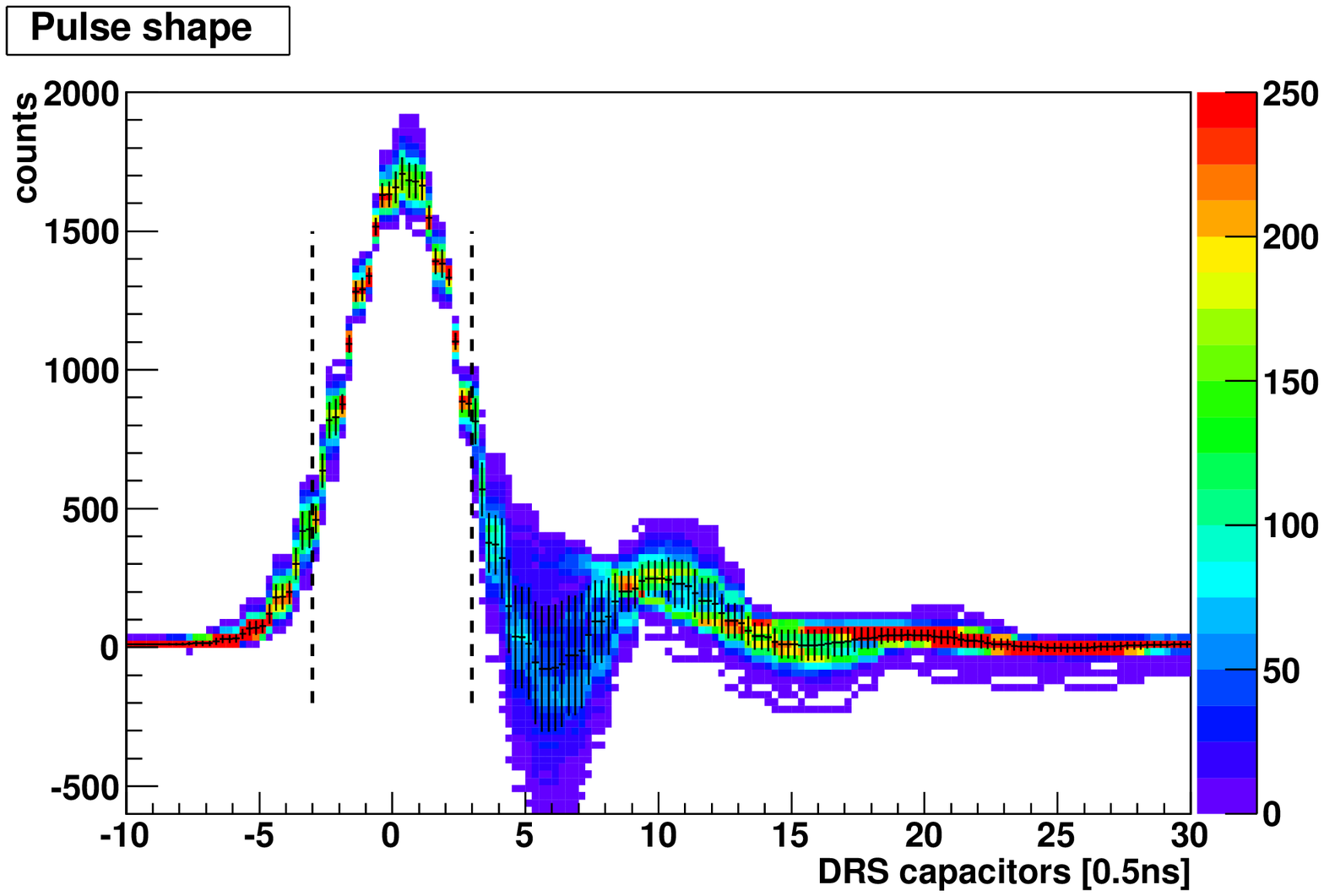}
\caption{
Spread of the calibration pulse shapes for the MAGIC~II telescope equipped with the DRS4 readout. 
Top panel: individual calibration pulses for a typical pixel.
Bottom panel: spread of average pulse shapes for all the channels. 
The color scale indicates the number of calibration events (top) or channels (bottom) and the black points show its profile.
Vertical dashed lines show the used signal extraction range for a window of 6 time samples. 
}
\label{fig:pulseshape}
\end{figure}

The shape of the waveform of a calibration laser pulse is shown in Fig.~\ref{fig:pulseshape}.
The main pulse is followed by a second smaller peak (overshoot) \citep{bpt13}.
In addition, the pulses in the DRS4 based readout are typically narrower than those from the DRS2, due to the bandwidth limitation of the latter.
The average pulse shape can be relatively well fitted with a sum of 3 Gaussian distributions (one of them with a negative amplitude), however the fit parameters vary from one channel to another.  
The overshoot has an amplitude of $\sim 18\%$ of the main peak and occurs between $3\,$ns and $5.5\,$ns after the peak of the main pulse.
In most of the channels there is an undershoot between the main peak and the overshoot.

The pulse shape of an individual channel (the top panel of Fig.~\ref{fig:pulseshape}) is very stable. 
However, the average pulse shapes are different for different readout channels. 
This effect depends on both the position of the channel inside the DRS4 chip and the high voltage (HV) applied to the PMT.

\begin{figure}
\includegraphics[width=9truecm]{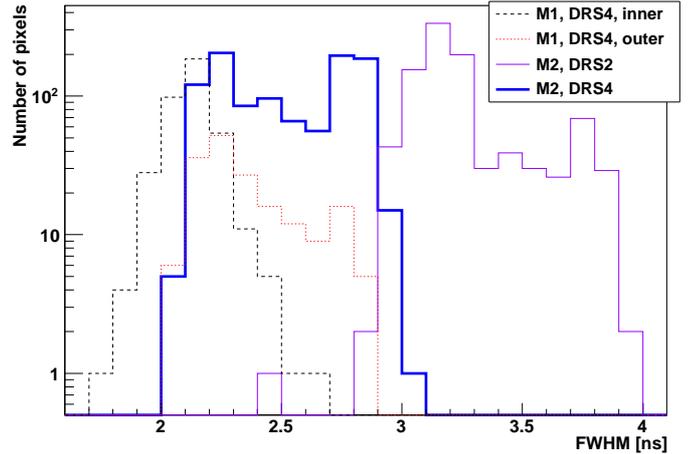}
\caption{
Distribution of the Full Width Half Maxima of a calibration pulse for all the channels for
inner (thin black dashed line) and outer (thin red dotted line) pixels of the MAGIC~I camera with DRS4 readout,
the MAGIC~II camera with DRS2 readout (thin violet solid line) and 
the MAGIC~II camera with DRS4 readout (thick blue solid line).}
\label{fig:fwhm}
\end{figure}

The distributions of the Full Width Half Maxima (FWHM) of the calibration pulses for various camera/readout setups used in MAGIC are shown in Fig.~\ref{fig:fwhm}.
The smallest FWHM ($\sim2.1\,$ns) is obtained for inner pixels of the MAGIC~I camera, equipped with the DRS4 readout. 
The outer, bigger pixels show, mainly due to lower HV, a slower response and also a broader distribution of the FWHM. 
MAGIC~II PMTs are operated at a lower voltage and do not have a fixed voltage between the photocatode and the first dynode, thus their response is in general slower, resulting in a broader FWHM. 
In addition, the type of the PMTs (produced by Electron Tubes Enterprises) used in MAGIC~I camera is intrinsically faster than the one used in MAGIC~II.
The second component at $\sim2.8\,$ns for MAGIC~II (see thick blue line in Fig.~\ref{fig:fwhm}) is composed largely of pixels whose HV had to be reduced in the HV flat-fielding procedure (see Section~\ref{sec:flatfield}) because the gain in the PMT or rest of electronic chain was especially high.

Since, contrary to the previous 0.3~GSamples/s readout used in the MAGIC~I telescope until 2007 (cf.~\citet{magic_fadc}), the waveform is sufficiently fine-sampled, we can use a relatively simple algorithm to extract the signal, the so-called ``sliding window''. 
We integrate the waveform in a fixed number of time samples (typically 6), but we scan the integration range over the entire readout window (of 80 time samples, i.e. 40~ns) in order to obtain the highest value of the integral. 
Since the size of the extraction window is smaller than the range of the overshoot, the extractor is not influenced by the secondary peaks.   
The sliding window provides also an estimator of the arrival time of the pulse as the mean time sample weighted with the signals in individual time samples. 
The signal extraction is done offline.

\begin{figure}[t]
\includegraphics[width=9truecm]{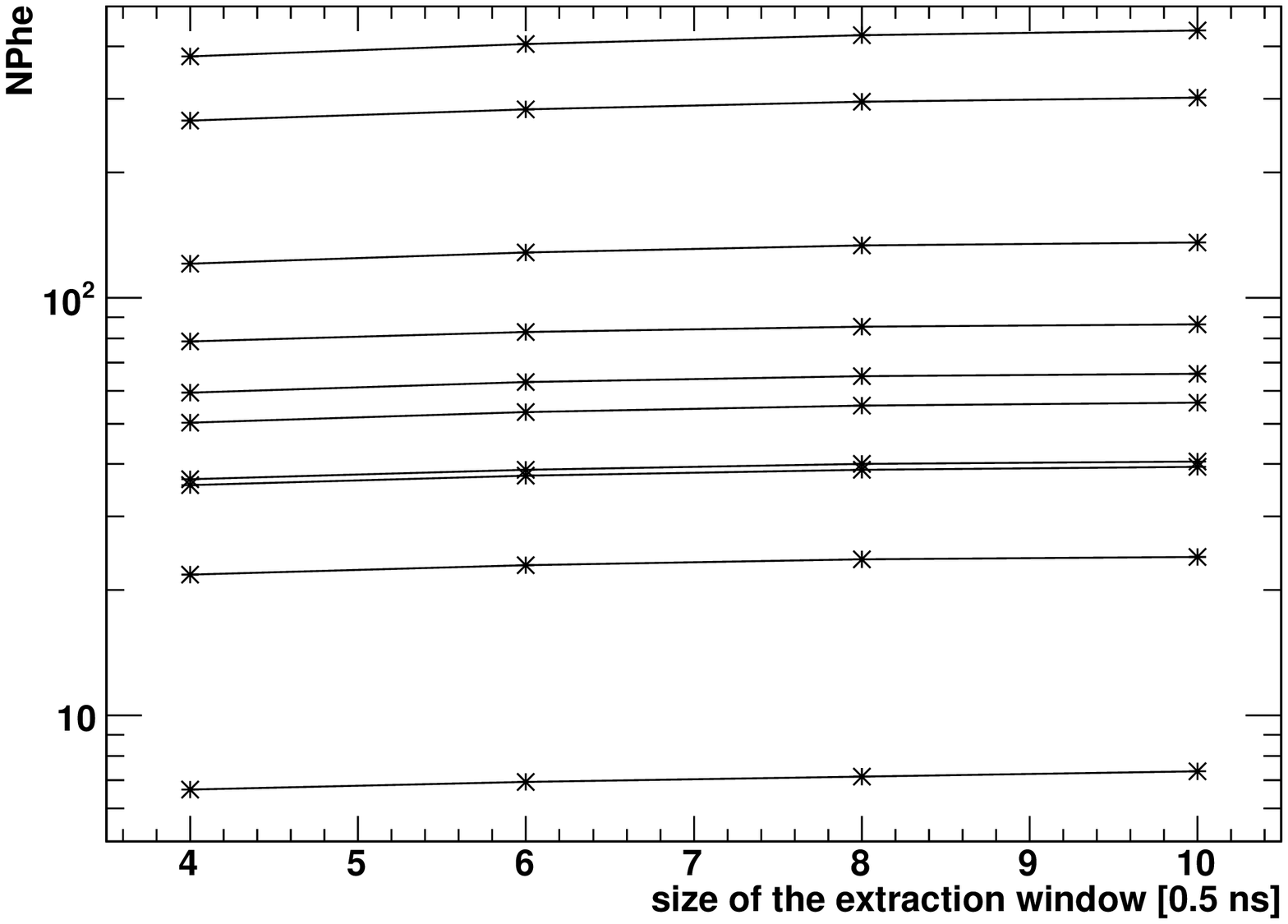}
\caption{
Extraction of pulses with different extractors:
number of photoelectrons reconstructed from calibration pulses with different sizes of the extraction window.
Individual lines show different light intensities.
The statistical uncertainty on the estimation of the number of photoelectrons is much smaller than the size of the markers.
}
\label{fig:extraction}
\end{figure}

In Fig.~\ref{fig:extraction} we show how the mean reconstructed number of photoelectrons depends on the size of the extraction window.
Having a small integration window is desirable as it maximizes the signal to noise ratio, but the size of the window should be large enough to cover most of the pulse.
The difference in the reconstructed number of photoelectrons is $\lesssim 5\%$ for the largest extraction window, covering the entire pulse, compared to the smaller, 6~time samples wide extractor window.
For example, for a typical light intensity at which the calibration is performed the reconstructed number of photoelectrons is 82.8 phe for a 6-sample window and 86.6 phe for a 10-sample window.

\subsection{High voltage flat-fielding}\label{sec:flatfield}
In order to achieve a good reconstruction even for faint images we want to have an optimised signal to noise ratio, also for small signals.
The gains of the PMTs can be set by the applied high voltage (HV).
Different HV settings vary the ratio between the noise from the light of the night sky (LONS) and the electronic noise.
However, reducing the influence of the electronic noise at a cost of increased HV will decrease the dynamic range. 
In addition we want to obtain a homogeneous response of all pixels to a signal that flashes the camera with a homogeneous density of photons.
This procedure is called the HV flat-fielding and it is performed with the calibration laser light pulses at the wavelength of 355$\,$nm. 
The HV were tuned to correspond to about 7000 counts in readout for a photon density of the order of 300 per PMT. 
This photon density corresponds on average to about 90 photoelectrons per pixel.

\subsection{Time calibration}
Due to small differences in the length of the optical fibers, in electronic paths and mainly the transit times of the electrons inside the PMT, (mainly caused by different high voltage settings), a synchronous short light pulse illuminating the camera will not be recorded in identical DRS samples for all readout channels. 
We found individual channel-to-channel delays of a few nanoseconds. 
Moreover, both the DRS2 and DRS4 chips exhibit an additional delay of typically $1\,$ns (up to $4\,$ns), depending on the absolute location of the signal pulse in the domino ring.

\begin{figure}[t]
\includegraphics[width=9truecm]{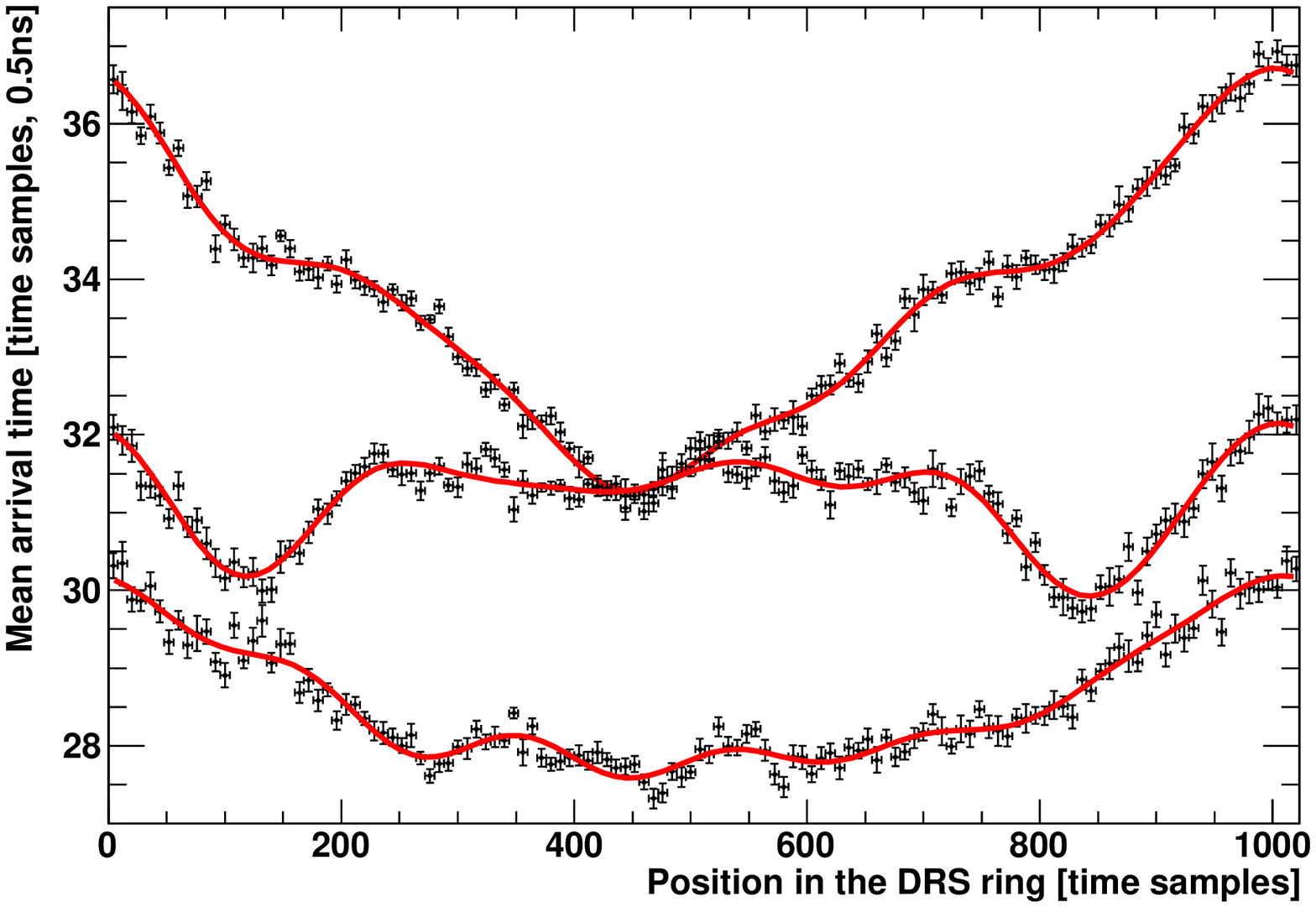}\\
\includegraphics[width=9truecm]{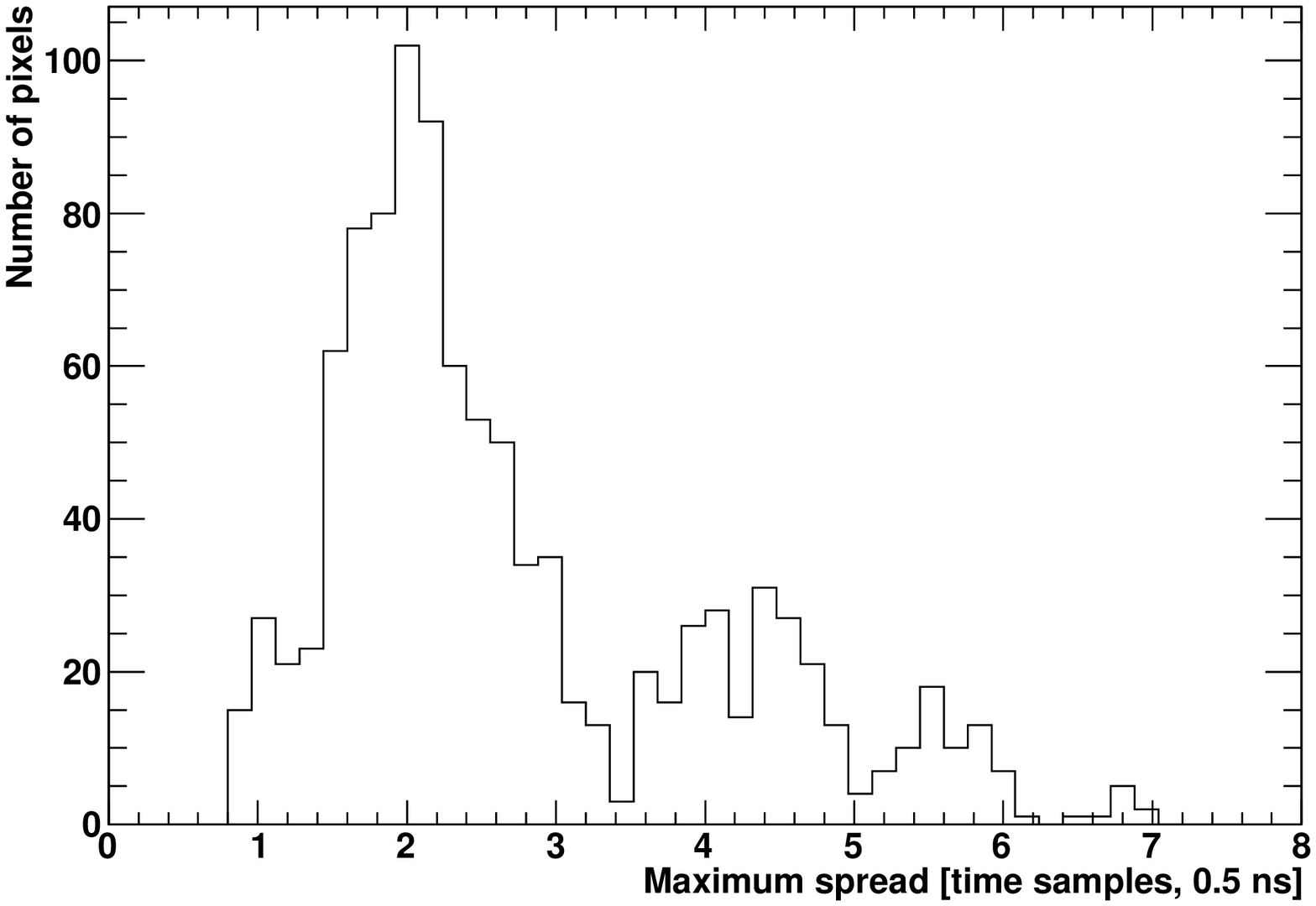}
\caption{Calibration of the time response of the DRS chip.
Top panel: mean arrival time as the function of the position in the DRS ring (data points) for 3 example channels together with their Fourier series expansion (lines).
Bottom panel: distribution of the maximum spreads of the DRS time delay for all channels.
}
\label{fig:drs_timecalib}
\end{figure}

As long as both the electronic path (optical fiber, receiver, DRS etc) and the HV of the pixel are not changed, the differences can be calibrated with the use of calibration pulses. 
The computed mean arrival time of a pulse in a channel is a rather complicated function of the time sample position in the ring (see Fig.~\ref{fig:drs_timecalib}) and differs from one DRS chip to another.
For each channel we expand this function into a Fourier series to obtain the correction function. 
Most of the DRS chips have a moderate jitter with a total spread of $\sim$1 ns, however in the distribution of all the DRS chips, one can see a long tail.
This calibration of the arrival time is done offline.

\begin{figure}[t]
\includegraphics[width=9truecm]{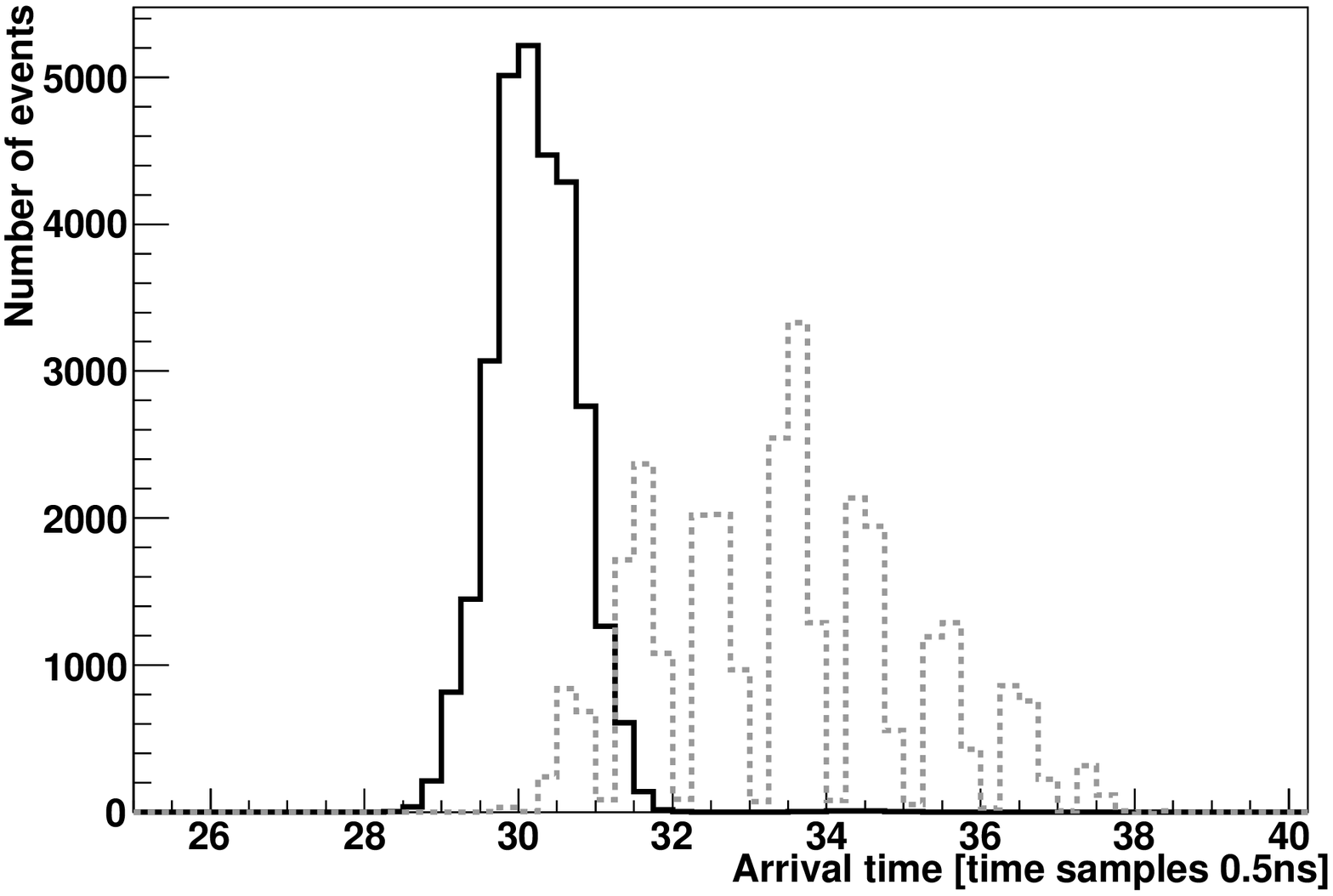}\\
\includegraphics[width=9truecm]{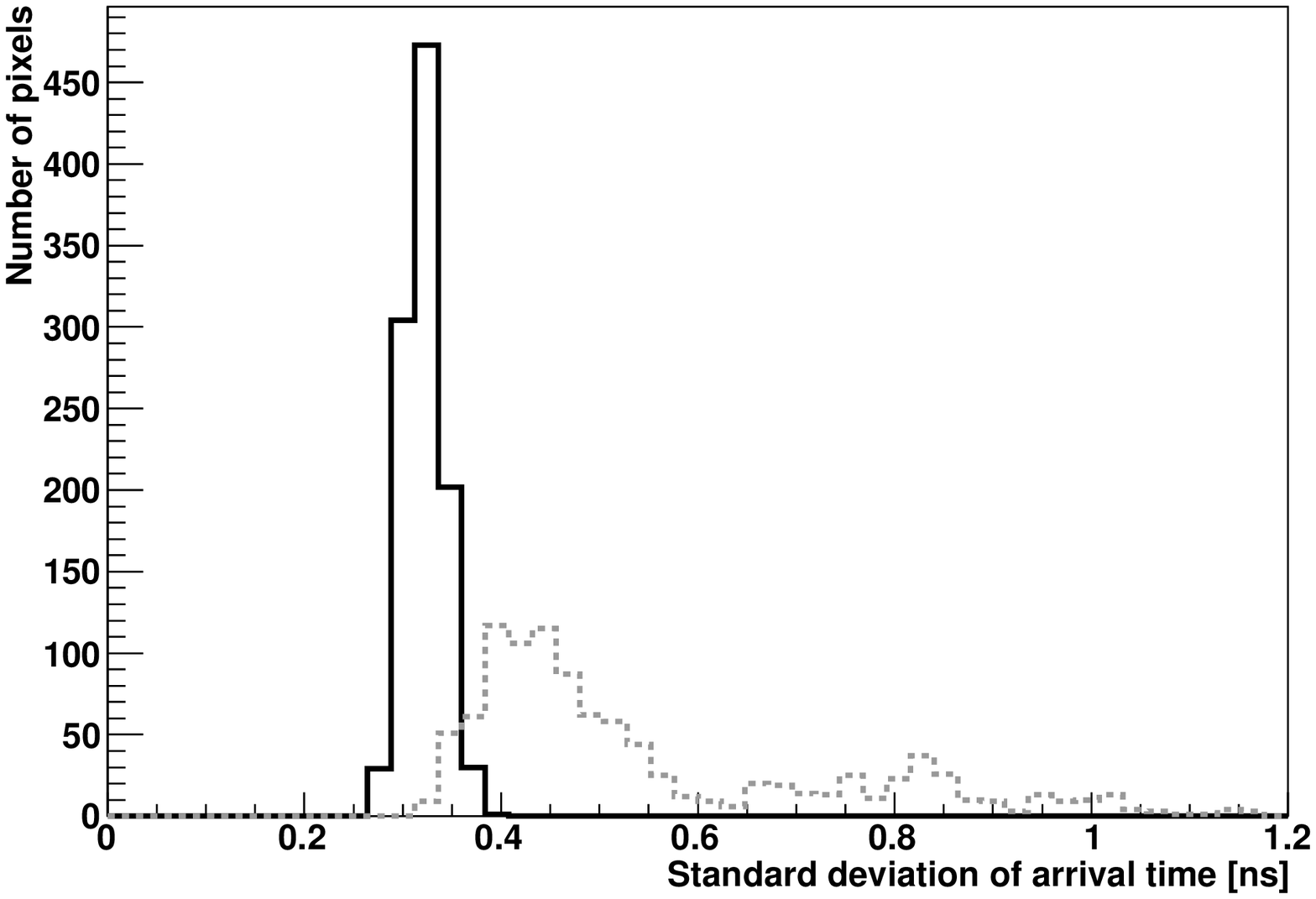}
\caption{
Arrival times of the calibration pulses before (dotted gray curves) and after (black solid lines) the calibration of the DRS time response.
Top panel: distribution of arrival times for a single pixel, 
bottom panel: distribution of the standard deviations of arrival times for all pixels.}
\label{fig:drs_timecalib2}
\end{figure}

In the top panel of Fig.~\ref{fig:drs_timecalib2}, we show the distribution of the arrival times of the calibration pulses, before and after such a calibration. 
The distribution of the uncalibrated arrival times shows multiple peaks due to 
the discrete values that the pulse integration boundaries can take. 
Since the spread of the DRS time delay is larger than one time sample, this structure is not visible anymore after the time calibration. 

The time calibration significantly reduces the standard deviation of this distribution for a $\sim 90\,$phe light pulse down to $0.32\,$ns (see the bottom panel of Fig.~\ref{fig:drs_timecalib2}).
Note, that this includes also the global jitter of the calibration pulse trigger (which was measured to be a flat distribution with a total width of $\sim 0.7\,$ns, i.e. corresponding to an RMS of about $0.2\,$ns). 

\begin{figure}[t]
\includegraphics[width=9truecm]{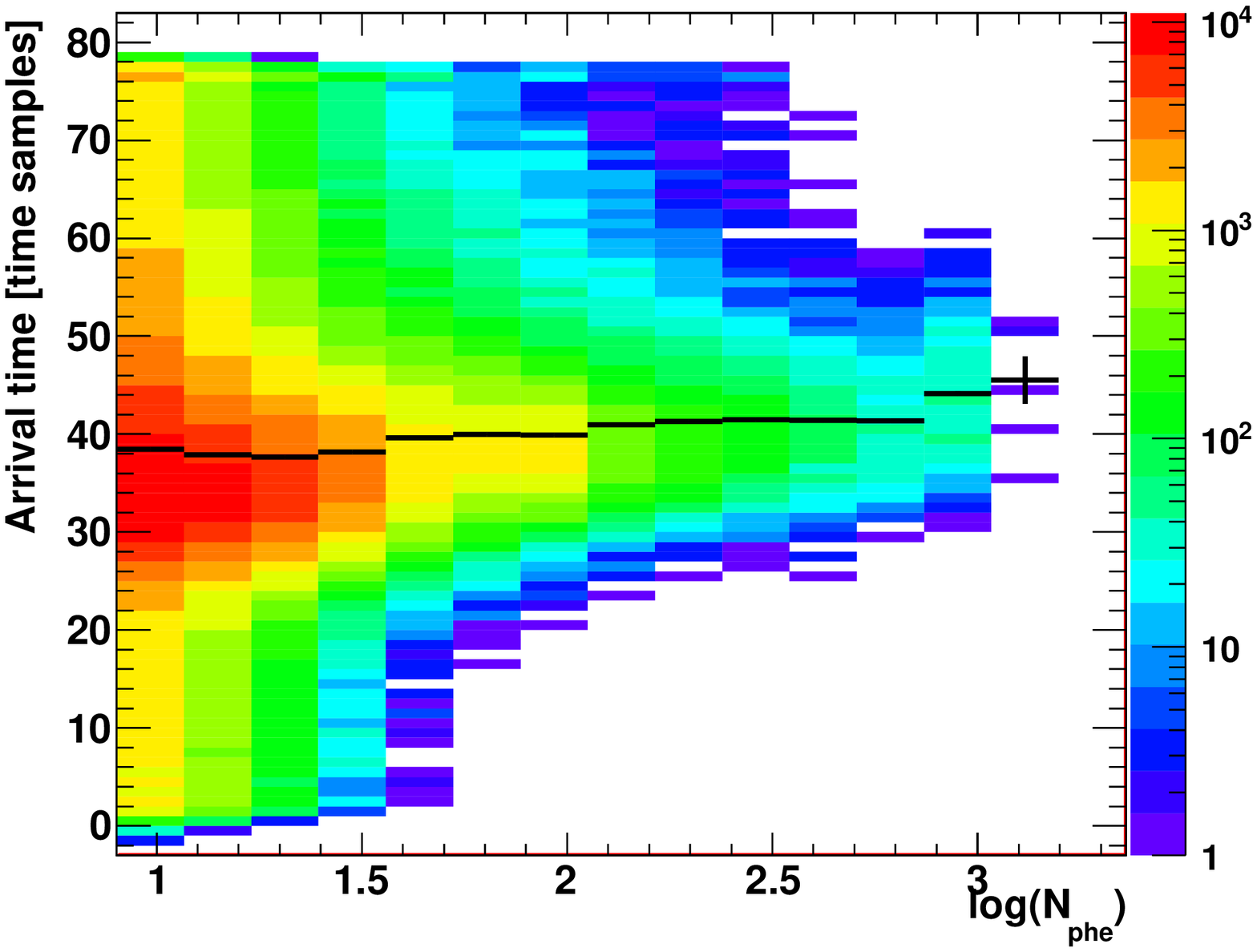}
\caption{
Two dimensional distribution of the arrival times of signals in individual channels from events triggered by cosmic-ray showers as a function of the signal strength for MAGIC~II. 
The black line shows the profile of the distribution.
}
\label{fig:tvsq}
\end{figure}

In Fig.~\ref{fig:tvsq} we show the two-dimensional distribution of the reconstructed arrival times and signals from events triggered by cosmic-ray showers. 
Each entry is for one pixel in a cosmic event, in total $\sim10^4$ random cosmic events have been used for this plot.
As the shower images usually cover only a small part of the camera, the low signal part of the distribution (for individual pixel signals of $\sim10\,$phe) is dominated by the tail of the baseline and noise fluctuations as well as afterpulses
\footnote{Afterpulses are rare signals with large amplitudes caused by an ion accelerated back to the photocatode of a PMT.}. 
The timing of the trigger signal is adjusted such that most of the pixels with signals belonging to cosmic-ray showers are centered around the middle of the readout window.
However, very large cosmic-ray showers with large impact parameters produce long images in the cameras with a long tail in the time distribution. 

\section{Performance of the readout}
In this section we report on the study of the most important performance parameters connected with the signal extraction of MAGIC data.
They have an influence on high-level performance parameters of the telescopes, such as the energy threshold, the sensitivity and the angular resolution \citep{magic_stereo}.

\subsection{Pedestal RMS and bias}\label{sec:ped_rms_bias}
\begin{figure}[pt]
\includegraphics[width=9truecm]{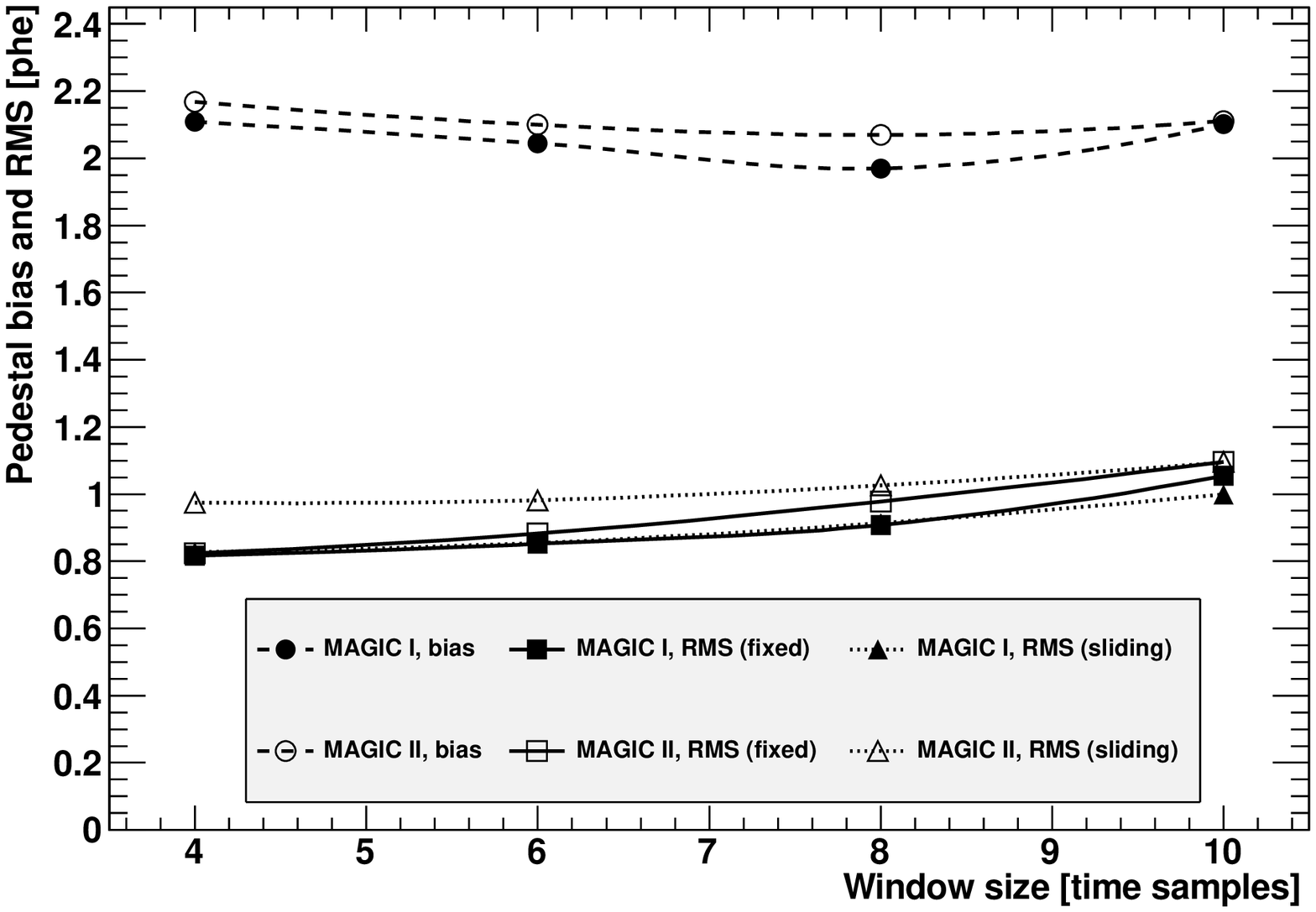}
\caption{
RMS and bias of the pedestal extraction as a function of different sizes of the integration windows for the observations of a dark patch of a sky using DRS4 readout. 
The square symbols and the solid line show the RMS for a fixed window extraction, while the triangles and the dotted lines show the RMS for a sliding window. 
The bias for a sliding window is shown with the circles and the dashed lines. 
The sliding window is allowed to search in the whole readout window of 40ns. 
Full symbols: the MAGIC~I camera (only smaller pixels), empty symbols: the MAGIC~II camera.
}
\label{fig:bias_rms}
\end{figure}

The size of the extraction window has to be large enough to accommodate most of the signal.
On the other hand, a too large window will spoil the performance of the extractor for low amplitude signals. 
We study the bias and the RMS (defined as the second central moment of the distribution) of the extractor. 
For the calculation of the bias, we allow the extractor window to search throughout the whole readout window of $40\,$ns.
We calculate the RMS both with a biased extractor (the so-called ``sliding window'') and with an unbiased one (the so-called ``fixed window'').
In the case of the fixed window we integrate a given number of samples starting at a random position in the total readout window, while for the sliding window we search for an integration window that provides the largest charge.
In the case of small (or lack of) signals the fluctuations of noise determine the position of the extraction window. 
Therefore the corresponding value of the pedestal RMS is dominated by the ``signal'' positions found by the sliding window extractor.
On the other hand if the signal is sufficiently larger than the fluctuations of the noise, the extraction window is determined by the pulse. 
The proper quantity to consider for the additional reconstructed pulse fluctuation is rather approximated by the pedestal RMS from a fixed window extractor.
Note that we estimate the RMS by directly computing the square root of variance of the estimated signal and not by fitting a Gaussian to the distribution of the obtained signals. 
This way, we take into account also non-Gaussian tails of the electronic noise, and the afterpulses generated by LONS photons. 
The sigma of the pure Gaussian part of the noise is $\sim20\%$ lower than the RMS. 
The results for both MAGIC cameras equipped with the DRS4 readout are shown in Fig.~\ref{fig:bias_rms}. 
Both the bias and the unbiased RMS of the pedestal are similar for both telescopes. 
The bias of the sliding window extractor is largely determined by the size of the total readout window, and only weakly depends on the integration window.
For the total window of 40ns the bias is nearly constant at a value of $2.1\,$phe.
On the other hand the RMS of the pedestal increases with the size of the window as more noise is integrated. 
Note that while the bias of the ``sliding window'' extractor is rather large, an additional precise information on pulse arrival times obtained from this extractor is required for constructing intelligent image cleaning algorithms \citep{magic_time}.

Assuming a rate of $0.13\,$phe/ns produced by the LONS from a dark extragalactic patch of sky~\citep{magic_mux}, we performed a toy MC simulation, computing the number of photoelectrons in the search window of $3\,$ns within the total time window of $40\,$ns.
We obtained that the mean value of such a bias is 1.5$\,$phe with AC coupling of the signal. 
Thus we conclude that the bias value of $\sim2.1\,$phe per $40\,$ns window seen in the data is mostly produced by the LONS.

\subsection{Non-Gaussian noise component}
Part of the noise computed in section~\ref{sec:ped_rms_bias} cannot be described by a Gaussian distribution. 
We estimate its contribution by computing the fraction of pedestal events with signals above a given threshold. 

\begin{figure}[t]
\includegraphics[width=9truecm]{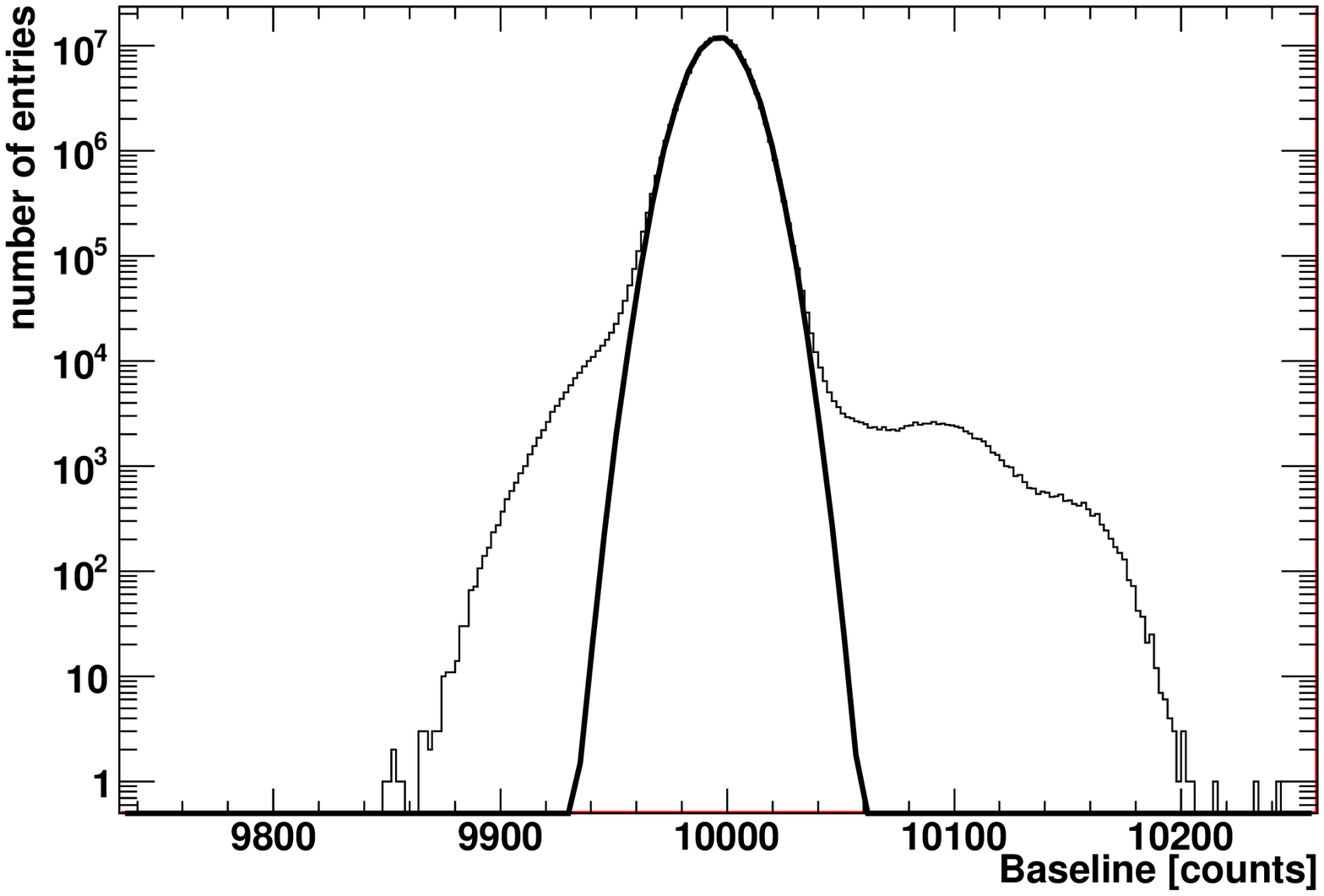}\\
\includegraphics[width=9truecm]{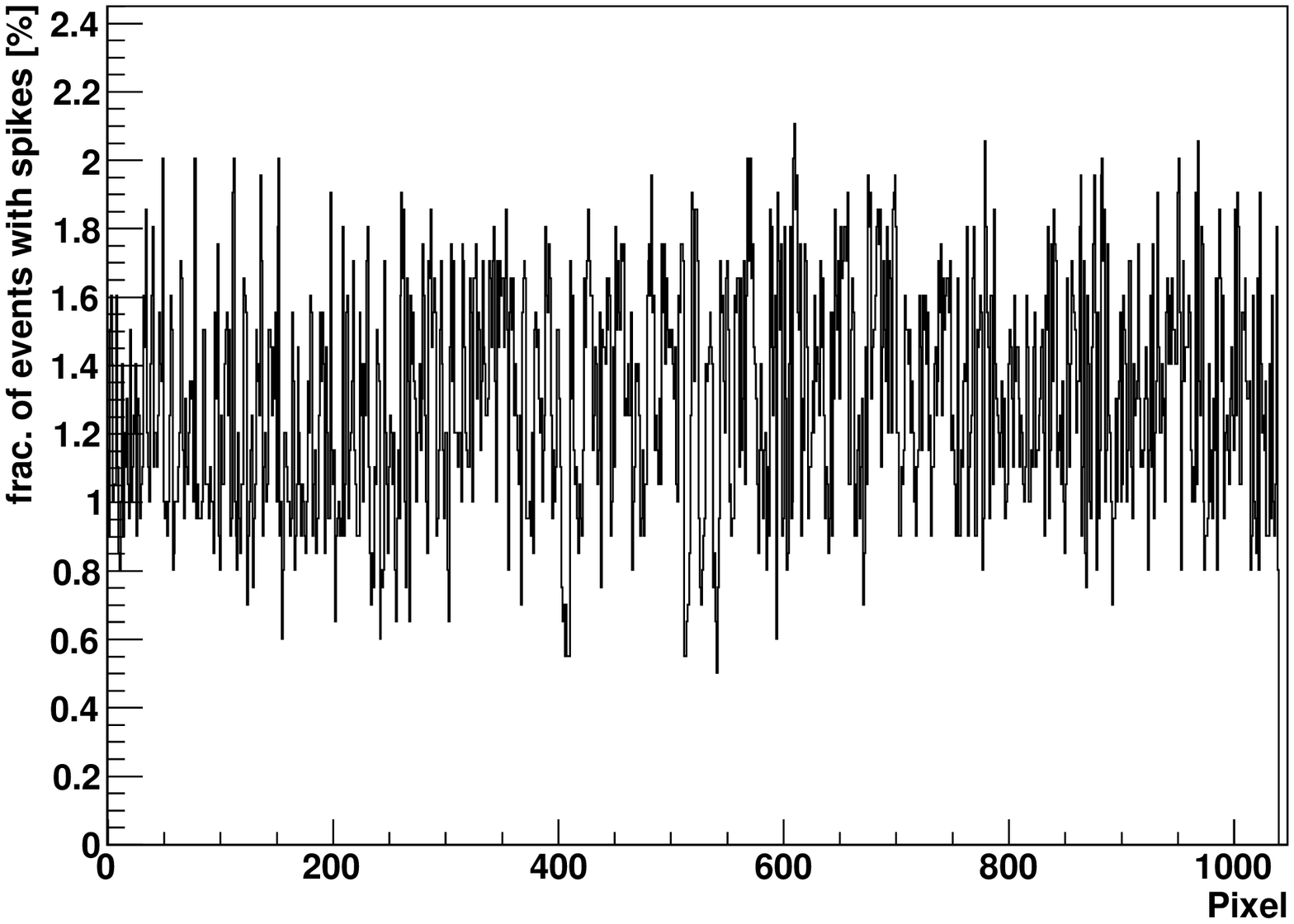}
\caption{
Top panel: distribution of the total electronic noise (closed camera run taken at the frequency of 300 Hz) of MAGIC~II with DRS4 readout (thin line) and the Gaussian fit (thick line), note logarithmic scale in Y-axis.
Bottom panel: fraction of events with signal above 100 counts for individual pixels. 
}
\label{fig:pedestal_hist}
\end{figure}

The distribution of the noise is shown in Fig.~\ref{fig:pedestal_hist}.
The deviations from the Gaussian distribution is at the level of a fraction of one per cent.
Analysis of a constant frequency pedestal run shows that about 1.2\% of events are affected by non-Gaussian electronic noise.
In those events there is an artificial signal (only due to noise fluctuations) greater than 100 counts in at least one capacitor.
Note that a single photoelectron produces on average a signal of an amplitude of $\sim30$ counts, and typically the integrated charge in a 6 sample extraction window is 90 counts). 

\subsection{Sources of noise}
The noise computed in section~\ref{sec:ped_rms_bias} can be either induced by the LONS or by the electronics. 
In order to find the dominant source of the noise in the MAGIC telescopes, we took a series of runs with random trigger in various hardware setups:
\begin{enumerate}
\item normal run with the open camera
\item run with the closed camera
\item run with the HV switched off
\item run with the camera powered down, i.e. the VCSELs switched off
\end{enumerate}
The last run allowed us to measure the noise induced by the DRS4 readout and the receivers (dominated by the DRS4 noise).
In order to check the noise produced in the camera (mostly generated by the VCSELs), one can subtract in quadrature the noise of the run 4 from the noise of the run 3. 
Similarly, the second run  minus the third run gives an estimation of the noise introduced by the HV, and the first run minus the second run tells the noise due to LONS. 
The results of those calculations are summarized in Table~\ref{tab:noise}
\begin{table}[t]
\centering
\begin{tabular}{c|c|c|c}
Source               & MAGIC~I            & MAGIC~II \\ \hline\hline
DRS4+receivers       & 0.45 (1.8) phe & 0.5 phe\\ \hline
VCSEL                & 0.4 (1.5) phe  & 0.3 phe\\ \hline
HV                   & 0.35 (2.1) phe & $\lesssim 0.1$ phe \\ \hline
LONS (extragalactic) & 0.40-0.55 (1.3-1.5) phe & 0.5-0.6 phe\\ \hline\hline
total &0.8-0.9 (3.4-3.5) phe  & 0.75-0.85 phe 
\end{tabular}
\caption{
Contribution to noise from different hardware components for inner MAGIC~I pixels (in parentheses values for outer pixels) and MAGIC~II, both equipped with the DRS4 readout.}
\label{tab:noise}
\end{table}

The ``natural'' noise connected with the LONS is of the order of the electronic noise. 
The dominant electronic noise comes from the readout in the control house, but in the case of the old MAGIC~I camera, the noise produced inside the camera was slightly larger. 
In the MAGIC~I camera a small noise of $\sim 0.35\,$phe is also induced by powering up the HV.
This noise might be caused by the proximity of the HV and LV lines in the old MAGIC~I camera.
The noise has been only measured shortly before the decommissioning of this camera.
Due to a different design such noise is not observed in the case of the MAGIC~II camera, and the noise value falls below accuracy of the measurement, $\lesssim 0.1$ phe.
As explained in Section~\ref{sec:flatfield}, both the inner and the outer pixels were tuned to have the same response (measured in readout counts) to the same photon density.
Since the outer pixels are 4 times larger in area, all the noise, which is not generated in the PMT itself (e.g. DRS4 noise), corresponds to 4 times larger noise, expressed in photoelectrons, for the outer pixels, compared with the inner ones. 
On the other hand, the LONS-induced noise of outer pixels is less than a factor 4 larger than in the case of the inner pixels. 
That noise is generated from a larger number of photoelectrons with smaller gain that in the inner pixels, which results in smaller relative fluctuations.

\begin{figure}
\includegraphics[width=9truecm]{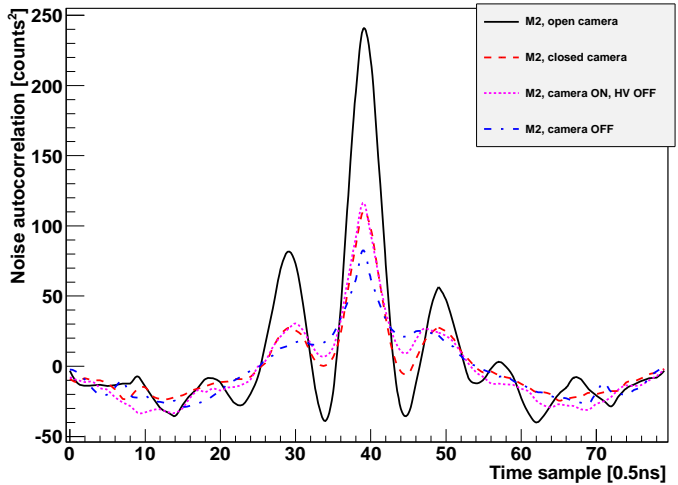}
\caption{
Noise autocorrelation matrix for sample number 40 for a typical channel in the MAGIC~II telescope operating with the DRS4 readout. 
The solid black line shows a normal run, the dashed red curve shows the closed camera run (only electronic noise), 
The magenta dotted line is the run with ramped down HV, and the blue dot-dashed curve shows a run with the camera powered down (only noise of the readout and the receivers).
}
\label{fig:noise_corr}
\end{figure}

Following the approach presented in \citet{magic_mux} we also investigate the noise autocorrelation function in MAGIC~II operating with the DRS~4 readout.
The noise autocorrelation function, $B_{ij}$, is defined as the correlation between the read-out samples $i$ and $j$:
\begin{equation}
B_{ij}=\langle b_i b_j\rangle - \langle b_i\rangle \langle b_j\rangle
\end{equation}
where $b_i$ and $b_j$ are the calibrated DRS4 sample content in time samples $i$ and $j$, and averaging is done over many events taken with random trigger.
In Fig.~\ref{fig:noise_corr} we show the noise autocorrelation function for a time sample $i=40$ (i.e. in the middle of the readout window) for various hardware setups. 
The main peak corresponds to the response of the readout to a single phe and/or noise. 
The secondary peaks are the result of the overshoot of the readout (compare with Fig.~\ref{fig:pulseshape}). 
The long wavy behaviour of combined DRS4 and receivers noise is a $40\,$MHz noise visible in some DRS4 chips (note however that the contribution of this noise to the total electronic noise is small).

\subsection{Dead-time}
\begin{figure}
\includegraphics[width=9truecm]{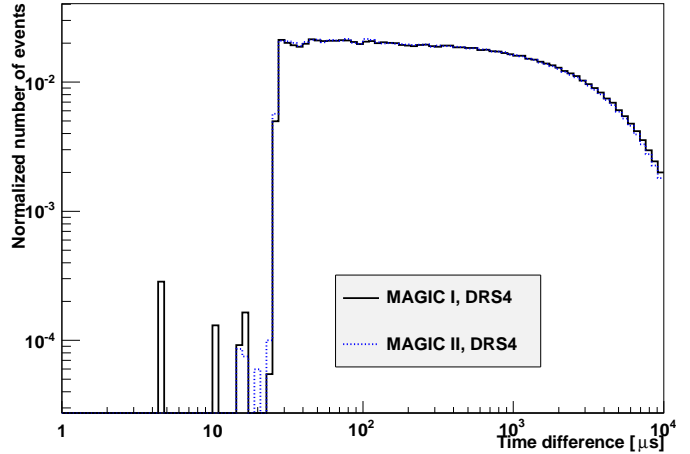}
\caption{
Distribution of the time lapses measured with the rubidium clock between consecutive events for the DRS4 readout in the MAGIC~I telescope (the solid black line) and the MAGIC~II telescope (the dotted blue line). 
}
\label{fig:deadtime}
\end{figure}

One of the advantages of the DRS4 with respect to the DRS2 chip is the possibility to read out only a fragment of the ring, the ``RoI''. 
In the RoI80 mode of operation we read 100 samples, but exclude on the fly the first and the last 10 samples, as they sometimes show higher noise than the 80 samples in between. 
This significantly decreases the dead-time per event from $500\,\mathrm{\mu s}$ for the DRS2 down to $27\,\mathrm{\mu s}$ for the DRS4.
The dead-time is constant and the same for all the DRS4 chips, thus it produces a sharp cut-off in the distribution of the time lapses between consecutive events (see Fig.~\ref{fig:deadtime}). 
The latency of the trigger systems used in the MAGIC telescopes ($\sim 35\,$ns  for the individual telescope trigger and $\sim 50\,$ns for the stereo trigger) is even smaller than this dead time.
As most of the events are recorded at the same time by both telescopes, the dead-time does not add up and the total dead-time of the present setup of MAGIC telescopes is also $27\,\mathrm{\mu s}$. 
A small fraction of events ($\sim 5\times 10^{-5}$) with apparent time lapses lower than the dead-time is due to rare jumps in the clock bits due to asynchronous latching of information. 
We have cross-checked the distribution of time lapses between the events using the internal clock of the readout and obtained a similar distribution.
Both the rubidium clock time-stamp and the DRS clock value are read using this method, thus the fraction of events with corrupted time lapses is similar.
But since the information from both clocks is supplied with different cables the events for which it happens are different.

The upgrade of the readout from DRS2 to DRS4 thus allowed us to reduce the total fraction of events lost due to the dead-time from $\sim 12\%$ down to a value of $0.7\%$ for a typical data taking rate of $250\,$Hz.

\subsection{Charge resolution}
\begin{figure*}
\includegraphics[width=6truecm]{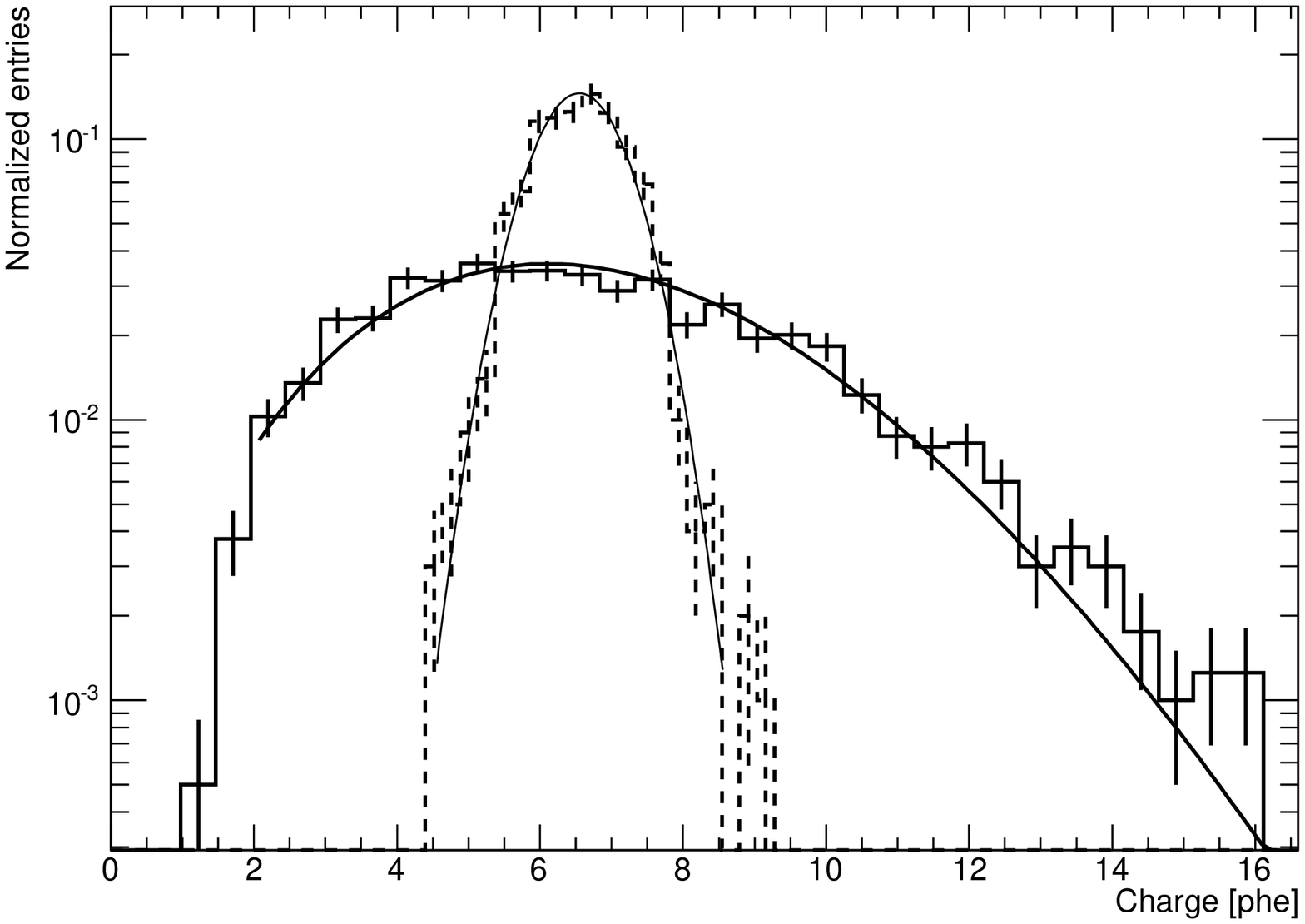}
\includegraphics[width=6truecm]{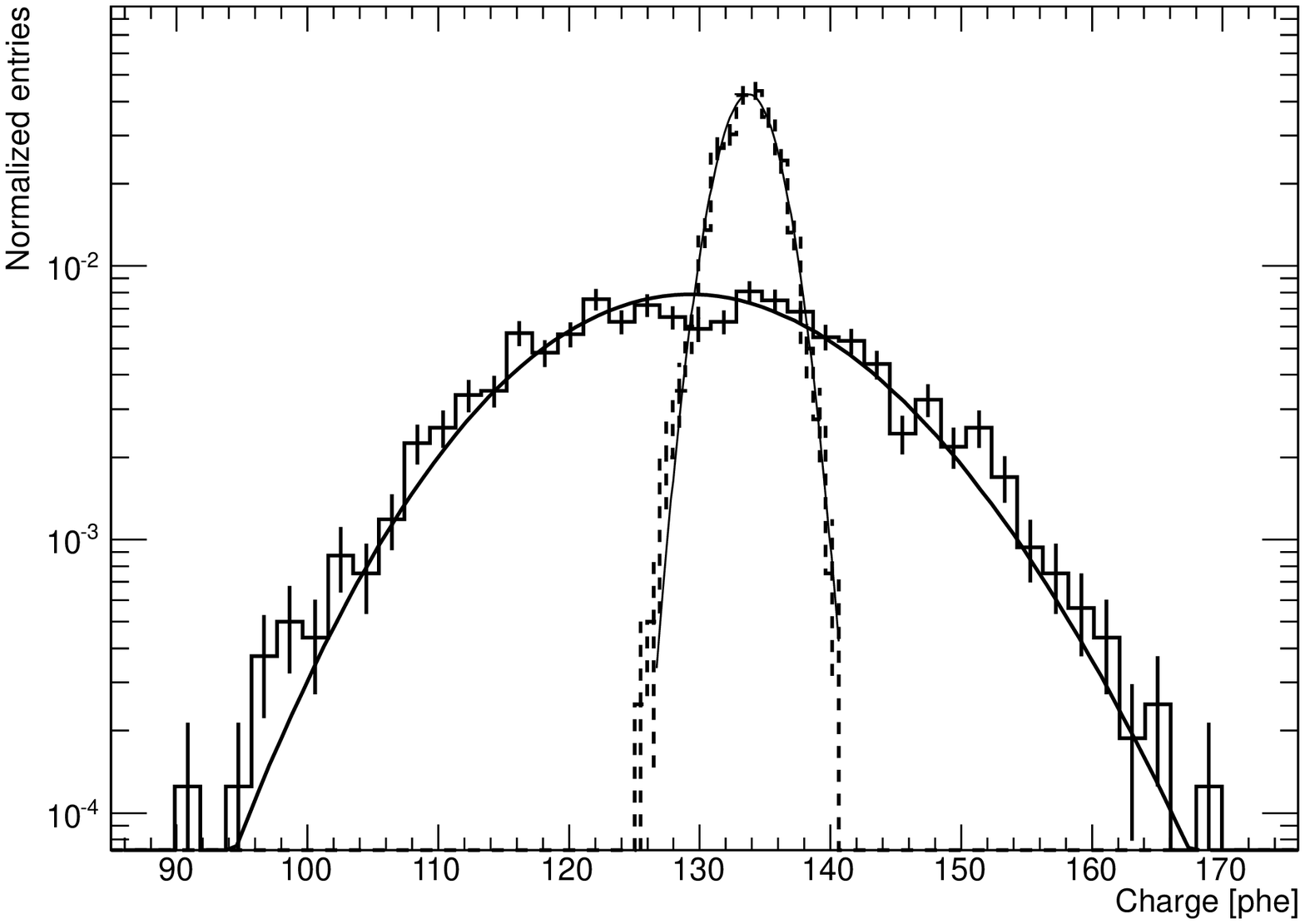}
\includegraphics[width=6truecm]{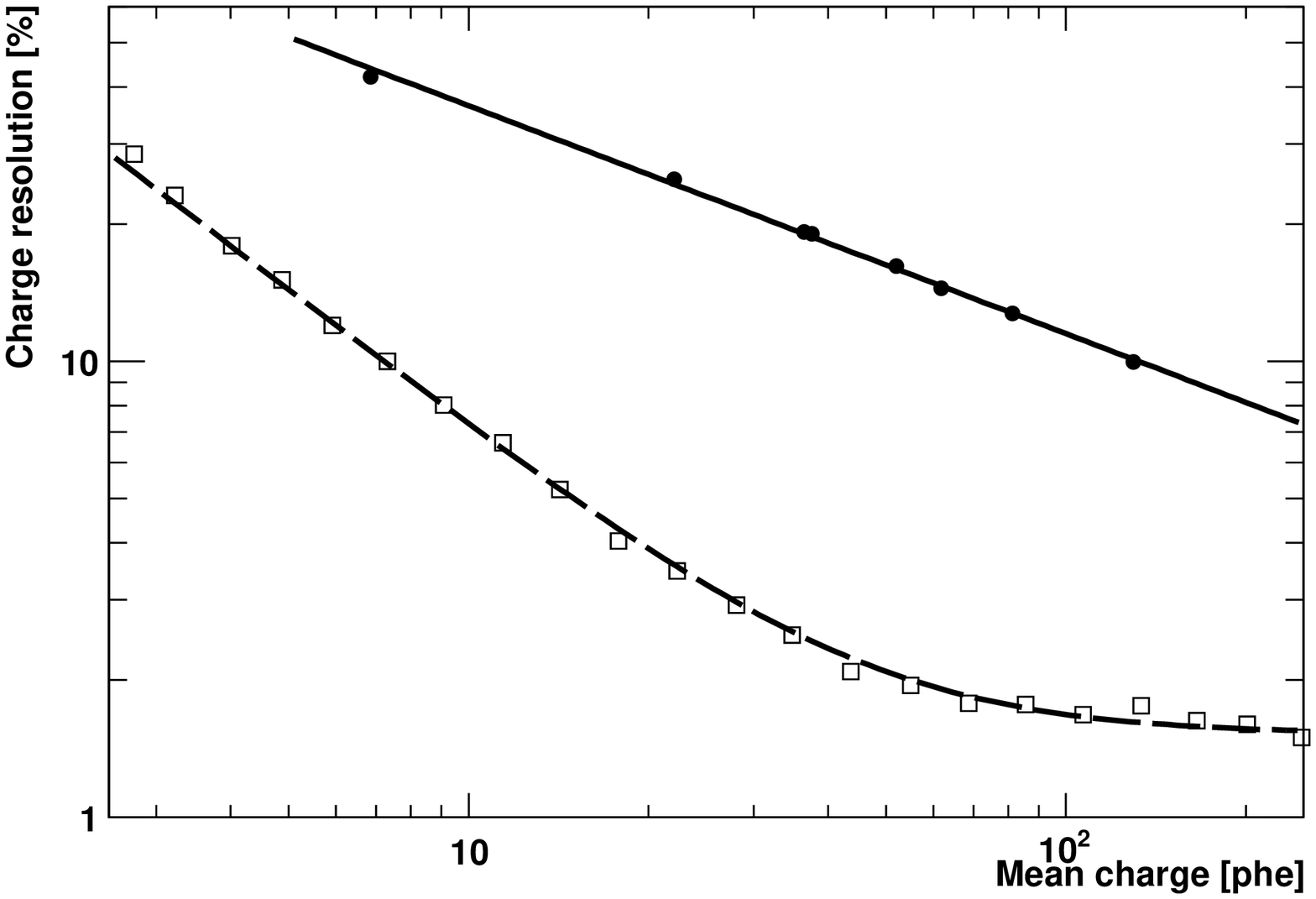}
\caption{
Distributions of the reconstructed signal for small (left panel) and large (middle panel) signals for the pulse injection (dashed lines) and for the light pulses (solid lines). 
Charge resolution (right panel) as a function of the (equivalent) number of photoelectrons for calibration light pulses (filled circles) and electric pulse injection (empty squares) for a typical channel equipped with a DRS4-based readout.
Statistical errors in the right panel plot are smaller than the size of the markers. 
}
\label{fig:chargeres}
\end{figure*}

We investigate the charge resolution as a function of the signal strength. 
We used both the calibration light pulses with different intensities, as well as electric pulses injected at the base of the PMT with various intensities. 
In Fig.~\ref{fig:chargeres} left and middle panels we show the distribution of the reconstructed charges for pulse injection and light pulses.
We apply the same conversion factors from the integrated readout counts to photoelectrons and calibrate pulse injection data in equivalent photoelectrons to allow for an easy comparison.
The distribution of the reconstructed charge for the pulse injection can be well described with a Gaussian.
On the contrary, the number of photoelectrons registered with a PMT follows Poissonian statistics (corrected for an assumed excess noise factor), and its distribution is much broader than for the pulse injection case. 
The distributions in the very low signals range ($\lesssim 2\,$phe) are affected by the bias of the extractor.

The charge resolution (defined as the RMS of the reconstructed charge distribution divided by its mean) is $\sim 10\%$ for pulse injection signals that resemble the ones produced by $7\,$phe in the PMTs and drops inversely proportionally to the signal strength. 
For very large signals it approaches a constant value of $1.5\%$. 
In turn the signal resolution for light pulses is strongly limitted by the Poissonian fluctuations of the number of generated photoelectrons. 
At a light intensity of $7\,$phe it is $\sim 40\%$ and drops with the square root of the number of photoelectrons. 
One should note that the pulse injection data, contrary to the light pulses, do not include the noise of the light of the night sky. 
Including it would spoil the charge resolution at the low and medium signals by 30-40\% (compare with Table~\ref{tab:noise}). 
However even including this effect the charge resolution is a factor of a few worse for light pulses than for electric pulses.
We conclude that the charge resolution of the DRS4-based readout cannot be fully exploited due to the large fluctuations of the number of photoelectrons registered in the PMT. 

\subsection{Time resolution \label{sec:timeresolution}}

One of the important signal processing performance parameters is the time resolution for signals of different light intensities. 
The time resolution curve can be parametrized by 3 parameters (see~\citet{magic_fadc} for details):
\begin{equation}
\Delta T = \sqrt{
\left(T_0 / \sqrt{N_{phe}}\right)^2 +
\left(T_1/ N_{phe}\right)^2 + T_2^2 }.
\label{eq:timeres}
\end{equation}

The $T_0$ parameter includes contributions of all Poissonian processes. 
In particular the intrinsic time spread of the photons and different travel times of individual photo-electrons produced at different places in the photocatode (and amplifying dynodes) contribute to $T_0$.
The $T_1$ parameter mostly depends on the pulse shape and the signal reconstruction resolution. 
The constant component, $T_2$, can be produced e.g. by a jitter of the clock of the readout or instrinsic time jitters of electronic components. 

In order to study the time resolution as a function of the signal strength, we took a series of data runs with calibration pulses of different laser light intensities. 
Using standard MAGIC simulation software, we also generated MC runs with different light intensities. 
Note that the intrinsic time spread of the laser pulse (FWHM = $1.1\,$ns, as measured in the lab, corresponding to a standard deviation of a Gaussian of $0.47\,$ns) will contribute quadratically to the $T_0$ parameter. 
Compared to that, the intrinsic time spread of photons produced in a gamma-ray shower in a single pixel is very similar and of the order of FWHM=$1\,$ns, or $\sigma\sim0.4$\,ns. 
However, as the laser light directly illuminates the camera (without being reflected from the mirror), the staggering of the mirrors of the MAGIC~I telescope
\footnote{the mirrors of MAGIC~I telescope were mounted in a chess-board pattern in two layers separated by $\sim6-8\,$cm} 
is not included in this study. 
The staggering of the mirrors will result in an arrival time distribution of individual photons from the showers convoluted with a bimodal distribution of the total spread of $\sim0.6\,$ns. 
We used toy MC simulations and determined that this is equivalent to adding in quadrature half of the time difference, i.e.~0.3\,ns, to the $T_0$ parameter of Eq.~\ref{eq:timeres}. 
The trigger of the calibration pulse has a small jitter (a flat distribution with the total spread of $\sim0.7\,$ns).
As this is a global trigger jitter, all the channels jitter uniformly, and there is no effect on the relative time differences between the signals in different channels. 
However, it is important to take this into account when checking the time response of a single channel. 
Therefore, for each event, we first calculate the mean arrival time from all channels, and subsequently the standard deviation of the distribution of the arrival time in a given pixel, minus the mean arrival time of all pixels. 
The time distributions could be fitted well with a single Gaussian for medium and large signals. 
For small signals of just a few photoelectrons in addition to the Gaussian peak there is a second flat component (which reflects the case where the signal extractor was confused by the residual noise or no photoelectron at all from the calibration light pulser arrived at the first dynode of the PMT), but the fitting range was adjusted to keep only the main Gaussian peak. 

\begin{figure}
\includegraphics[width=9truecm]{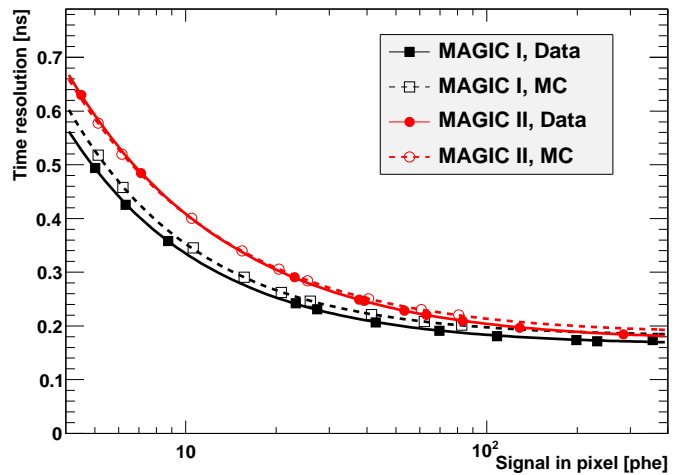}
\caption{
Time resolution (standard deviation of a Gaussian distribution) for signals of different strength in MAGIC~I (black squares, only inner pixels) and MAGIC~II (red circles).
Filled symbols (fitted with a solid line) show the data, and empty symbols (fitted with a dashed line) show the numbers obtained from MC simulations.
}
\label{fig:timeres}
\end{figure}

The results are shown in Fig.~\ref{fig:timeres}.
One can see that the MC simulations can well reproduce the behavior of the data.
The time resolution for large pulses ($\gtrsim 100\,$phe) approaches $T_2 = 0.17\pm0.01\,$ns.
For the low signals of about $5\,$phe it is still as good as $0.5-0.6\,$ns. 
Due to different PMT type and HV settings the transit time spread in MAGIC~I is lower than in MAGIC~II, resulting in a lower  $T_0$ coefficient ($T_0^{\rm M1}=0.8\pm0.1\,$ns for MAGIC~I, $T_0^{\rm M2}=1.1\pm0.1\,$ns for MAGIC~II).
Since both telescopes are equipped with the same readout and have similar noise, the $T_1=1.5\pm0.3\,$ns parameter is the same for both of them.
The error on the fit parameters given above has been estimated by performing a similar fit to each pixel separately and computing the RMS of the distribution of the obtained values.
Therefore they should be treated as systematic errors arising from differences between individual pixels/channels. 

Taking into account the staggering of the mirrors and the typical time spread of photons from gamma-ray showers (adding quadratically 0.3\,ns and 0.4\,ns and subtracting 0.47\,ns), 
we expect thus for MAGIC~I gamma-ray showers: 
\begin{equation}
\Delta T^{\rm M1} = \sqrt{
\left(0.8 / \sqrt{N_{phe}}\right)^2 +
\left(1.5 / N_{phe}\right)^2 + (0.17)^2 } \qquad ,
\end{equation}
and for MAGIC~II (adding quadratically 0.4\,ns and subtracting 0.47\,ns): 
\begin{equation}
\Delta T^{\rm M2} = \sqrt{
\left(1.1 / \sqrt{N_{phe}}\right)^2 +
\left(1.5 / N_{phe}\right)^2 + (0.17)^2 } \qquad .
\end{equation}

As one can see, the correction from calibration light pulses to gamma-ray showers reflected by the mirrors, does not change the results within a precision 100\,ps.
For the reference value of $10\,$phe the time resolution is equal to $\Delta T^{\rm M1}=0.34\,$ns and $\Delta T^{\rm M2}=0.42\,$ns, while for $100\,$phe $\Delta T^{\rm M1}=0.19\,$ns and $\Delta T^{\rm M2}=0.20\,$ns.

\subsection{Linearity}

The DRS2 readout, used previously in the MAGIC~II telescope, was highly non-linear. 
Moreover, the non-linear response of the system could vary from one capacitor to another, even in a single readout channel. 
The non-linearity could be calibrated by applying a set of constant voltages at the input of the DRS2 and comparing them with the resulting signals (uncalibrated counts) in different capacitors. 
\begin{figure}[tbp]
\includegraphics[width=9truecm]{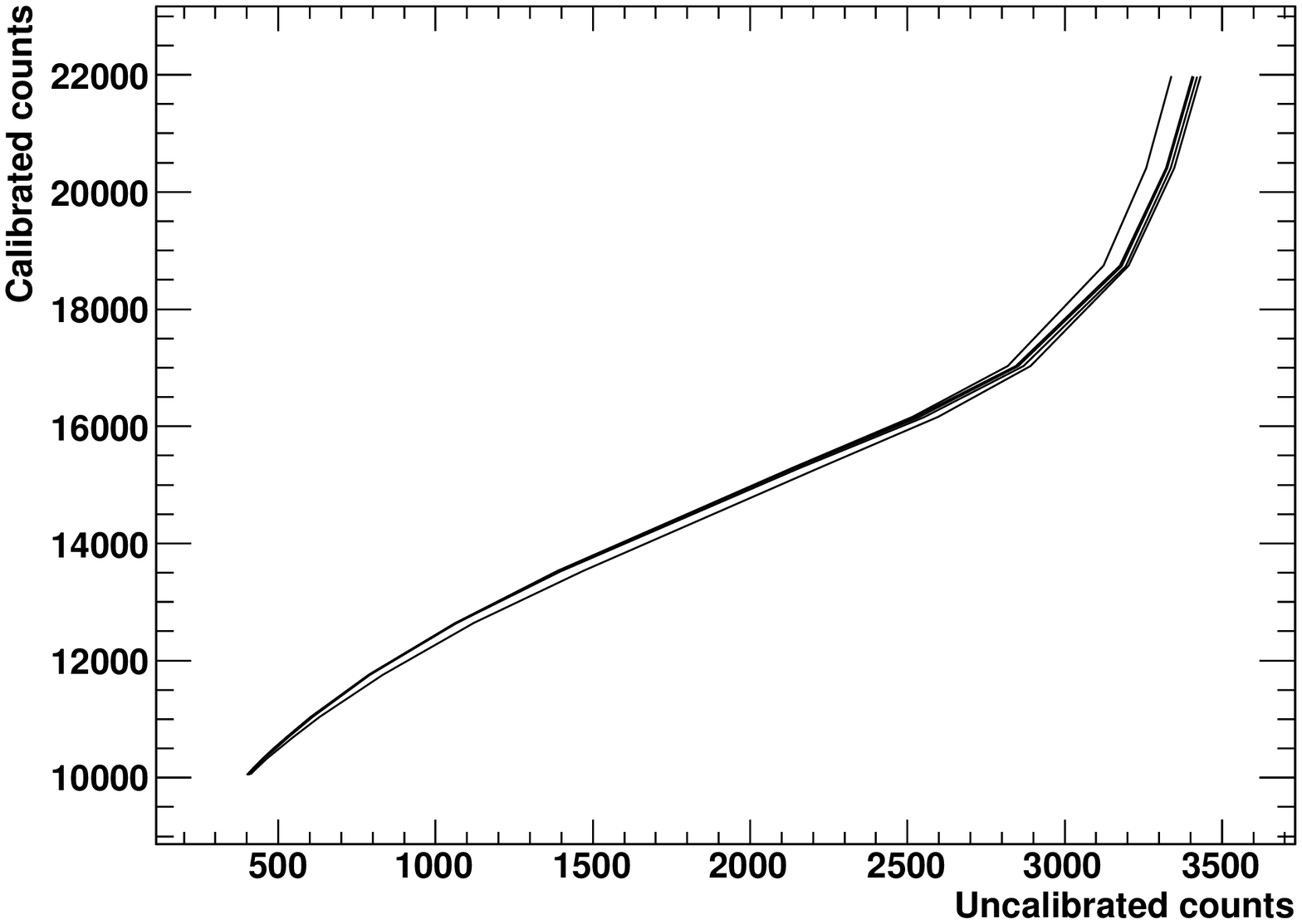}\\
\includegraphics[width=9truecm]{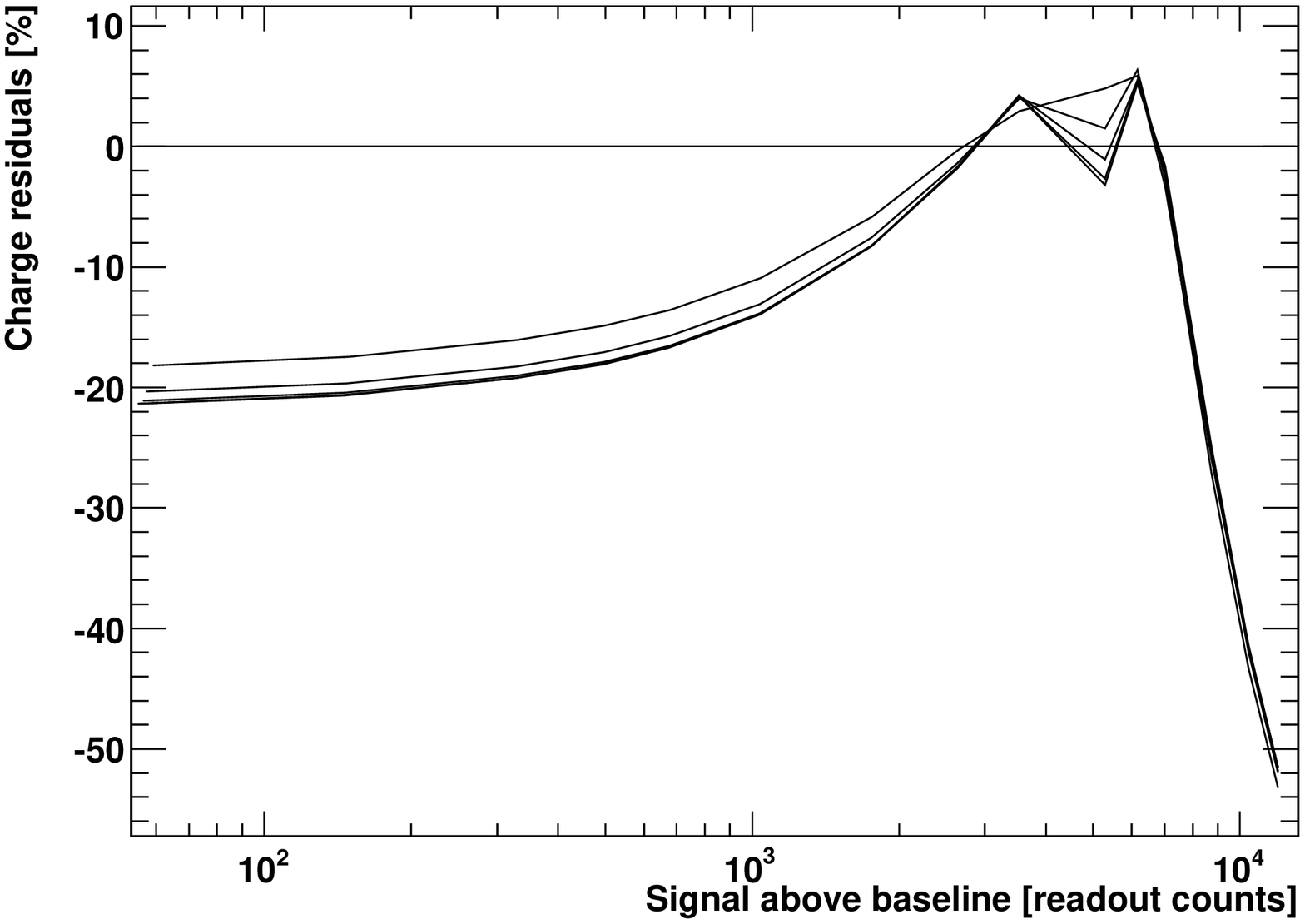}
\caption{
Top panel: linearization curves for 5 capacitors in a DRS2 channel. 
Bottom panel: deviation from linear behaviour, measured as a difference with respect to a linear fit, for those capacitors.
}
\label{fig:linDRS2}
\end{figure} 
The slope of the curve changes from about 6 for small signals down to 2.5 for medium signals (see Fig.~\ref{fig:linDRS2}). 
Very large signals ($\gtrsim$ 900~phe) saturated the DRS2 readout.
Knowing the linearity curves for each capacitor, the linearization could be done off-line. 

In contrary to that, the DRS4 readout has an excellent linear behaviour up to its saturation at the value of $\sim$13000 DRS4 counts in a single capacitor above the baseline. 
The apparent 3-5\% deviation from the linearity at the lowest charges are dominated by uncertainty of the input voltage used for the measurement.
Since a single photoelectron produces a signal with an amplitude of $\sim30\,$readout counts this would naively correspond to saturation starting at $400\,$phe. 
A large pulse which includes the intrinsic time spread of the calibration/cherenkov light flashes and the time spread in the PMT has an effective amplitude of $\sim18$ readout counts per photoelectron (see Section~\ref{sec:signal_processing}).
On top of that the integration window of 6 samples is less sensitive to the saturation than just the amplitude the pulse. 
Therefore, for a typical light pulse the saturation in DRS4 becomes important only above $\approx 750$~phe.
\begin{figure}[tbp]
\includegraphics[width=9truecm]{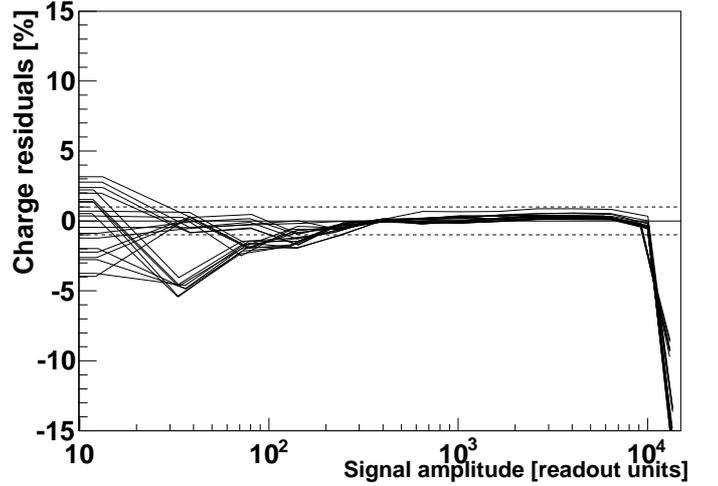}
\caption{
Deviation from linear behaviour, measured as a difference with respect to a linear fit divided by the input value, for 20 typical channels of the DRS4 readout. 
The horizontal dotted lines show non-linearity at the level of 1\%.
A single photoelectron has an amplitude of $\sim 30$ readout counts.
}
\label{fig:linDRS4}
\end{figure} 
The deviations from linearity are typically $\lesssim 1\%$ (see Fig.~\ref{fig:linDRS4}).

\subsection{Cross-talk}

\begin{figure*}[tbp]
\centering
\includegraphics[width=15truecm, height=8.5cm]{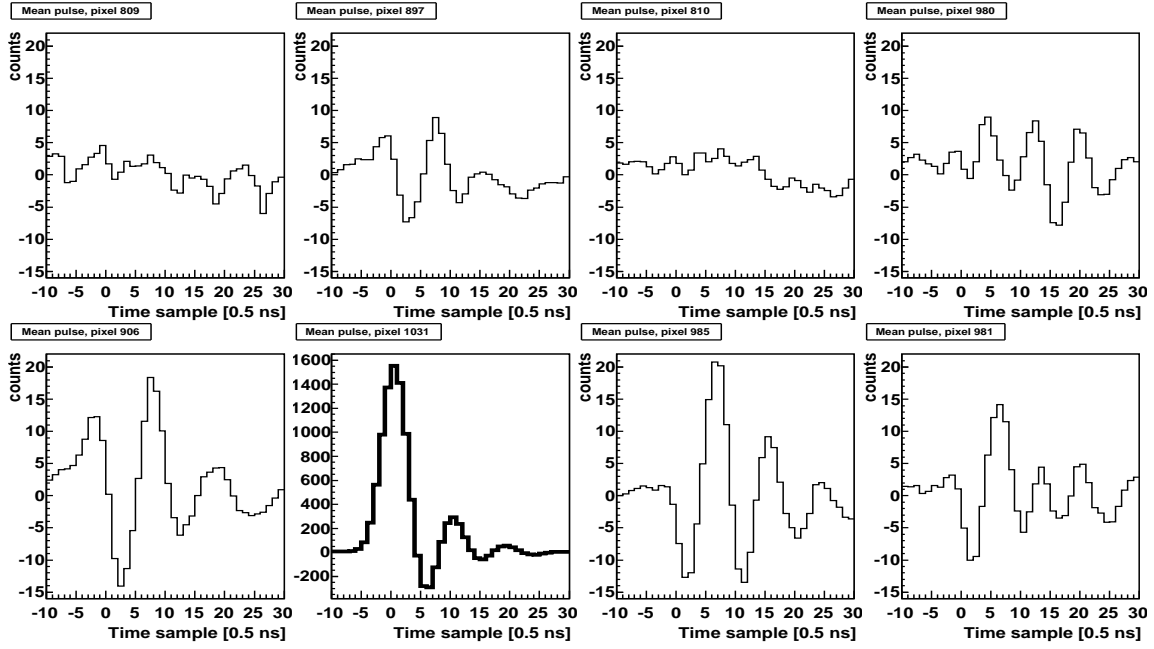}
\caption{
Shape of the cross-talk induced signals in DRS4 channels.
The original signal is injected in pixel 1031 (thick line, 6$^\mathrm{th}$~panel, showing a factor 80 larger scale than the other shown channels). 
}
\label{fig:xtalk_shape}
\end{figure*}

\begin{figure*}[t!]
\includegraphics[width=9truecm]{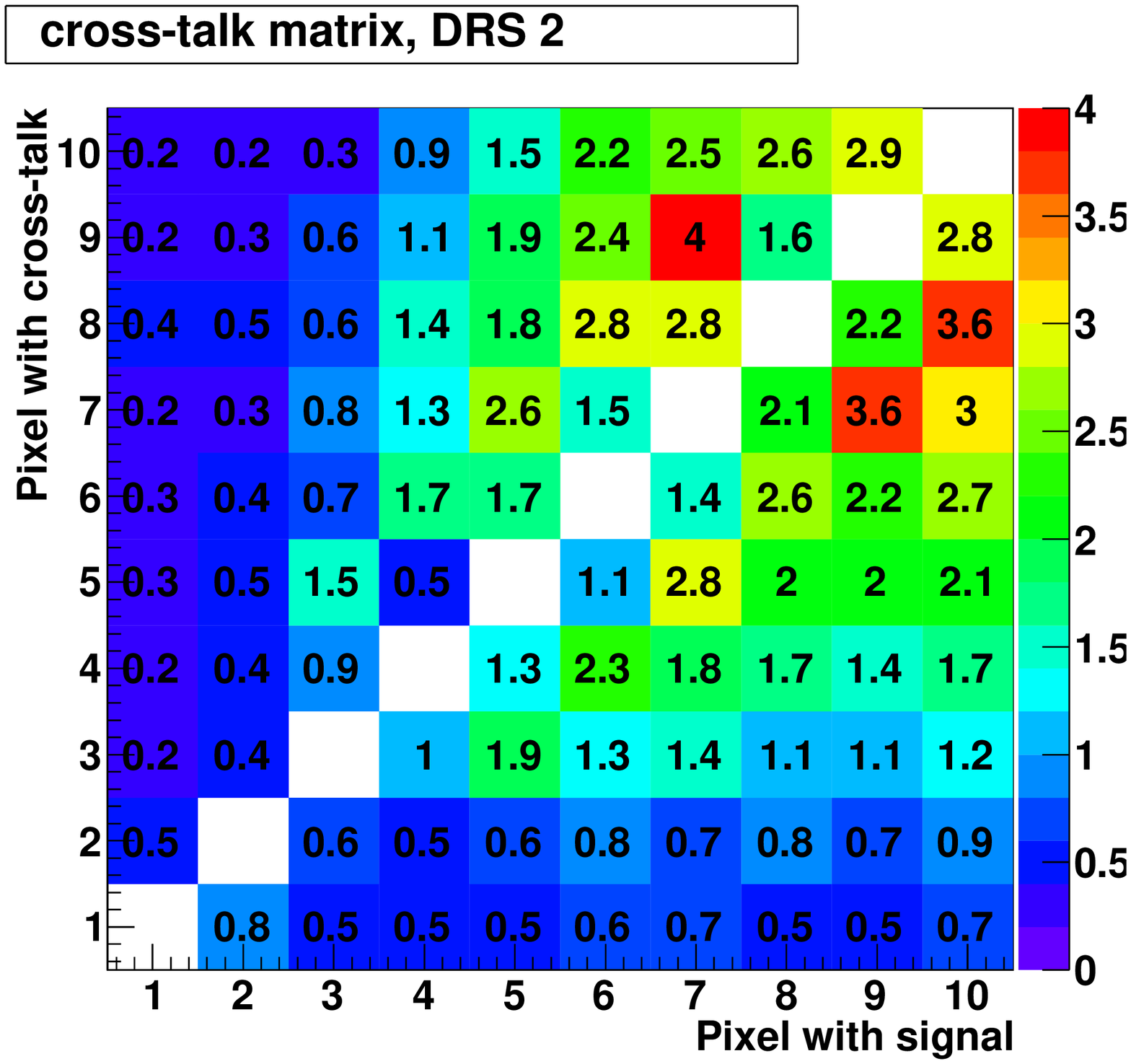}
\includegraphics[width=9truecm]{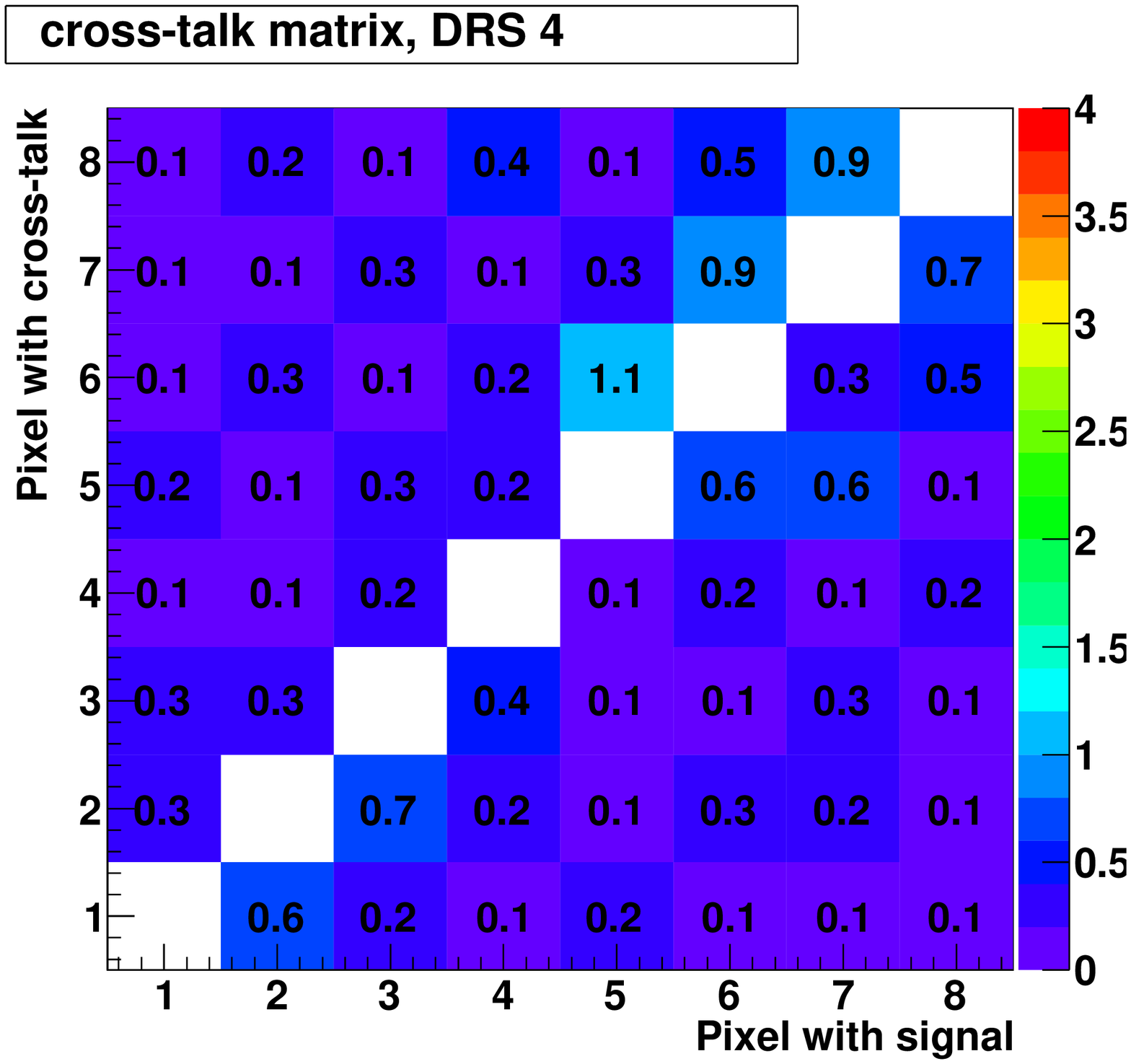}
\caption{
Cross-talk matrix for a DRS2 (the left panel) and a DRS4 (the right panel) chip.
The numbers and the color scale show which percentage of the original signal is induced in another channel due to cross-talk.
}
\label{fig:xtalk}
\end{figure*}

The DRS chips exhibit cross-talk between the channels of the same chip, 
the shapes of which are shown in Fig.~\ref{fig:xtalk_shape}.
The cross-talk matrices (giving the fraction of the original signal injected into another channel) are shown in Fig.~\ref{fig:xtalk} for typical DRS2 and DRS4 chips. 
The values of the cross-talk in Fig.~\ref{fig:xtalk} are computed for the signal extraction with integration windows normally used in the analysis of DRS2 and DRS4 data, i.e. 8 and 6 time samples respectively.

In the case of the DRS2 readout, a significant cross-talk (up to $\sim4\%$ of the orignal signal) was observed in most of the combinations of channels. 
As the cross-talk from different channels can pile up, the total cross-talk from 9 channels with the same signal can easily add up to $\sim10\%$ of the original signal. 
For large showers, signals of the order of 100~phe are common, and this would produce artificial signals of the order of 10~phe, which can easily exceed the thresholds set by the image cleaning. 
In order to correct for the cross-talk, we invert the cross-talk matrix and apply it to the reconstructed signals. 
This procedure was done only for the data taken with the DRS2 readout. 

In the case of DRS4, only a moderate cross-talk ($\sim 1\%$) is visible in the neighbouring channels. 
Farther channels in the same DRS4 chip have much lower cross-talk (of the order of $\sim 0.3\%$). 
On top of that, the total number of channels used in one DRS chip is lower (8 in the DRS4 chip compared to 10 in DRS2), thus pile-up of cross-talk is smaller.

\section{Conclusions and outlook}
We presented analysis methods used to extract and calibrate the charge and time information from individual pixels of the MAGIC telescopes. 
The upgrade of the readout of the MAGIC telescopes from DRS2 to DRS4 based has significantly improved the performance of the system.
An advanced pedestal subtraction procedure in DRS4 data results in a stable baseline. 
For typical observation conditions, the noise in both telescopes lies below one photo-electron, with comparable contributions from the electronic noise and the LONS. 
The pulses are being extracted with the ``sliding window'' extractor of a width of $3\,$ns.
The calibration of the time response of the DRS4 chip allows to obtain excellent time resolution of $0.2\,$ns for signals larger than a few tens of photoelectrons. 
Even for small signals of a few photoelectrons, the time resolution is still as good as $\sim0.5\,$ns.
The upgrade of the readout to DRS4 allowed to decrease the dead time from 12\% to a negligible fraction.
In contrary to the large non-linearity of the DRS2 readout, the linearity of the new system was proven to be very good up to the saturation at $\sim 13000$ counts above the baseline.
For the typical light pulse time spread the saturation occurs for signals above $750\,$phe resulting in a nearly three orders of magnitude dynamic range.
Also the cross-talk was reduced by a factor of a few to a value, which normally does not influence the data any more. 

All the here presented low-level performance has been proven to be sufficiently good for the application of a DRS4-based readout in IACTs. 
Moreover the MAGIC telescopes have been successfully using such a readout for the past 1.5 years. 
This makes the DRS4 a viable candidate for the signal digitization of the future CTA project. 
However, one should be aware that the excellent performance has been achieved only after complex software preprocessing of the raw data. 
In the case of highly integrated cameras and electronics of telescopes such preprocessing may become challenging. 

\section*{Acknowledgements}
{
\small
We would like to thank the MAGIC collaboration for allowing us the usage of hardware and software tools that were needed to perform this study. 
We would especially like to thank Rudolf Bock, Juan Cortina, David Fink, Razmik Mirzoyan, and David Paneque for many discussions and careful reading of the manuscript.
We also thank the two anonymous referees for their comments and suggestions which helped to improve the manuscript.
}

\end{document}